\documentclass[
notitlepage,
floatfix,
aps,prx,
reprint,
twocolumn,
superscriptaddress,
nofootinbib
]{revtex4-2}

\usepackage{graphicx}
\usepackage{tikz}
\usepackage{float}
\usepackage{xcolor}
\usepackage{hyperref}
\hypersetup{colorlinks,linkcolor={blue},citecolor={blue},urlcolor={blue}} 
\usepackage{enumerate}
\usepackage{siunitx}
\usepackage{graphicx}% Include figure files
\usepackage{dcolumn}% Align table columns on decimal point
\usepackage{bm}% bold math
\usepackage{physics}
\usepackage{mathtools}
\usepackage{xcolor}
\usepackage{booktabs}
\usepackage{multirow}
\usepackage[version=4]{mhchem}
\usepackage{amsthm}
\usepackage{amsfonts}
\usepackage{hyperref}
\usepackage{cleveref}
\usepackage{braket}
\usepackage{float}
\usepackage{tikz}

\newcommand{\bal}[1]{
\begin{align} #1 \end{align}
}

\newcommand{\baln}[1]{
\begin{align*} #1 \end{align*}
}

\def\lpar{\left(}
\def\rpar{\right)}
\def\lbra{\left[}
\def\rbra{\right]}
\def\lkey{\left\{}
\def\rkey{\right\}}

\def\H{\text{H}}
\def\AH{\text{AH}}
\def\Hsb{H_\text{SB}}
\def\Hdb{H_\text{DB}}

\def\sb{\text{SB}}
\def\db{\text{DB}}
\def\fb{\text{fb}}

\crefname{equation}{}{}
\Crefname{equation}{}{}
\crefmultiformat{equation}{#2(#1)#3}{–#2(#1)#3}{–#2(#1)#3}{–#2(#1)#3}
\creflabelformat{equation}{#2(#1)#3}

\renewcommand{\tr}[1]{\text{Tr}\left( #1 \right)}
\newcommand{\expec}[1]{\mathbb{E} \left( #1 \right)}
\renewcommand{\braket}[1]{\left\langle #1 \right\rangle}

\def\cC{ {\cal C} }

\def\cD{ {\cal D} }

\def\cT{ {\cal T} }

\def\cO{ {\cal O} }

\def\rhs{\text{r.h.s. }}
\def\hc{\text{h.c.}}

\def\real{\mathbb{R}}
\def\complex{\mathbb{C}}
\def\goto{\rightarrow}
\def\epsilon{\varepsilon}

\newcommand{\cL}{{\cal L}}
\newcommand{\be}{\begin{equation}}
\newcommand{\ee}{\end{equation}}

\newcommand{\bigo}[1]{{\cal O}\left({#1}\right)}

\renewcommand{\pmat}[1]{\begin{pmatrix} #1 \end{pmatrix}}

\newtheorem{theorem}{Theorem}
\newtheorem{coro}{Corollary}
\newtheorem{prop}{Proposition}

\begin{document}

\title{Hamiltonian and double-bracket flow formulations of quantum measurements}

\author{Aar\'on Villanueva}
\email{aaronv@science.ru.nl}
\affiliation{Faculty of Science, Radboud University, Heyendaalseweg 135, 6525 AJ Nijmegen, The Netherlands}
\affiliation{
Theoretical Division (T-4), Los Alamos National Laboratory, Los Alamos, New Mexico 87545, USA}
\author{Luis Pedro Garc\'ia-Pintos}
%\email{lpgp@lanl.gov}
\affiliation{
Theoretical Division (T-4), Los Alamos National Laboratory, Los Alamos, New Mexico 87545, USA}

\begin{abstract}
We introduce a framework that unifies quantum measurement dynamics, Hamiltonian dynamics, and double-bracket gradient flows. We do so by providing explicit expressions for stochastic Hamiltonians that produce state dynamics identical to those that happen during continuous quantum measurements. 
When such dynamical processes are integrated over sufficiently long time intervals, they yield the same results and statistics as during wavefunction collapse. That is, wavefunction collapse can be interpreted as coarse-grained (stochastic) Hamiltonian dynamics. 
Alternatively, wavefunction collapse can be interpreted as double-bracket gradient flows determined by derivatives of (stochastic) potentials defined in terms of observables with direct physical interpretations. The gradient flows minimize the variance of the monitored observable. 
Our derivations hold for general monitoring described by non-Hermitian jump processes. 
We show that such reinterpretations of measurement dynamics facilitate the design of feedback processes. In particular, we introduce feedback processes that yield deterministic double-bracket flow equations that prepare ground states of a target Hamiltonian, and state-agnostic feedback processes for state preparation.
We apply the latter for entanglement stabilization of a two-qubit system considering a setup with imperfect measurements and feedback delay.
We conclude by re-interpreting feedback processes as gradient flows with tilted fixed points.
\end{abstract}

\maketitle

%================================================================================

\section{Introduction}
Traditional interpretations of quantum theory divide dynamics into two distinct classes: Hamiltonian processes and measurement processes~\cite{nielsen2010quantum, griffiths2018introduction}.  
In the former, the state $\psi_t$ of isolated quantum systems evolves unitarily, governed by Schr\"odinger's equation $d \psi_t/dt \nobreak=\nobreak -i H \psi_t$. In such a regime, the system's Hamiltonian $H$ completely determines its dynamics.

Hamiltonian dynamics are in sharp contrast to what occurs during quantum measurements. In the latter, the collapse postulate says that, upon measuring an observable $A = \sum_j a_j \Pi_j$ with eigenvalues $a_j$ and eigenprojectors $\Pi_j$, the outcome $a_j$ is obtained with probabilities $\psi_t^\dag \Pi_j \psi_t$. The collapse postulate also states that, immediately after observing such an outcome, the system's state is $\Pi_j \psi_t/\sqrt{\psi_t^\dag \Pi_j \psi_t}$~\cite{nielsen2010quantum, griffiths2018introduction}. 
Open system dynamics due to interactions with an environment, or stochastic master equations arising from conditioning on measurement outcomes, can be derived from a combination of the two aforementioned regimes. 

Continuously monitored quantum systems provide the natural theoretical and experimental bridge between these two regimes~\cite{WisemanMilburn1993, wisemanQUANTUMMEASUREMENTCONTROL, Jacobs2014}.
In practice, measurements on quantum hardware are rarely instantaneous: when a quantum system is coupled weakly and continuously to a meter---be it a microwave cavity, an optical mode, or an atomic gas---the joint system–meter dynamics, upon conditioning on the output of real-time measurement outcomes, yields a stochastic evolution of the system state referred to as a quantum trajectory~\cite{DalibardCastinMolmer1992, Carmichael1993, WisemanDiosi2001, GambettaWiseman2003}.
This framework has been validated across a range of experimental platforms, including dispersive readout of ultracold atomic gases, single-shot observation of individual superconducting qubit trajectories, and heterodyne detection of fluorescence
~\cite{DalibardCastinMolmer1992, murch2008observation, murch2013observing, Hatridge2013, Weber2014}.
Beyond their foundational significance, continuously monitored systems are relevant to quantum feedback control~\cite{Doherty1999, Feedback2, Feedback3, Feedback4, Feedback5, Feedback6, Feedback7, Feedback8, Feedback9} and quantum error correction and  error mitigation~\cite{AhnDohertyLandahl2002, SarovarMilburn2005, PazSilva2014}, where real-time processing of measurement outcomes enables active steering, stabilization, and protection of fragile quantum states.
Understanding the mathematical structure underlying monitored dynamics is therefore not only a conceptual question, but a prerequisite for the systematic design of feedback protocols in emerging quantum technologies.

In this letter, we show that despite their distinct physical origins, measurement processes can be mathematically cast as Hamiltonian processes, which, in turn, can be interpreted as gradient flows. 
In Section~\ref{sec:MeasurementHamiltonian}, we show that arbitrary pure-state dynamics can be expressed in terms of single- and double-bracket Hamiltonian processes. We derive explicit expressions for stochastic Hamiltonians that enforce dynamics equivalent to measurement processes. Such Hamiltonians, which depend on the measured observable, the random measurement outcomes, and the system’s state, reveal that measurement dynamics involve a double-bracket form contribution interpretable as gradient flows, as shown in Section~\ref{sec:DoubleBracket}. In Section~\ref{sec:Feedback}, we leverage such reformulation of measurement dynamics to design novel Hamiltonian-based techniques for quantum feedback control that modify the minimum of the gradient flow.

%================================================================================

\vspace{7pt}
\section{Dynamics of continuously monitored quantum systems}
Continuously monitored quantum systems~\cite{barchielliQuantumTrajectoriesMeasurements2009} that are subject to a single noise channel characterized by a jump operator $C$ can be described by the following stochastic master equation (SME) in Itô form:
\bal{
\label{eq:sme_general_ito}
    d\rho_t = \cL(\rho_t) dt + \left\{C-\braket{A}_t, \rho_t \right\}_c dW_t\,,
}
where $\rho_t$ is the density operator.
The right-hand side involves the Lindblad super-operator $\cL(\rho_t) \coloneqq -i[H_0, \rho_t] + \cD[C](\rho_t)$, where $H_0$ is the system's Hamiltonian.
The dissipation part is $\cD[C](\rho_t) \coloneqq \gamma \lpar C\rho_tC^\dagger - \frac{1}{2} \lkey C^\dagger C, \rho_t \rkey \rpar$ with damping rate $\gamma > 0$, which we assume to be time-independent.
In the diffusion term, we define $\{X, Y\}_c \coloneqq XY + Y^\dagger X^\dagger$ for arbitrary operators $X, Y$.
The quantity $\braket{A}_t = \tr{A \rho_t}$ is the quantum expectation value at time $t$ of the Hermitian part of $C$ given by
\bal{\label{eq:A}
A \coloneqq (C+ C^\dagger)/2 \,.
}
The infinitesimal increment $dW_t$ denotes a Wiener (Brownian) noise term with zero mean, $\expec{dW_t} = 0$, whose variance over a time interval $dt$ is given by $\expec{dW_t^2} = \gamma\,dt$.
We will often omit the expectation symbol and write $dW_t^2 = \gamma dt$.

The state $\bar{\rho}(t) \coloneqq \expec{\rho_t}$ obtained by averaging over all noise realizations~\cite{barchielliQuantumTrajectoriesMeasurements2009} satisfies the Lindblad equation~\cite{GoriniKossakowskiSudarshan1976,Lindblad1976}:
\baln{
    \frac{d\bar{\rho}}{dt} = \cL(\bar{\rho})\,.
}

We assume from now on that $\rho_t$ describes a pure quantum state, that is, $\rho_t = \psi_t \psi_t^\dagger$ where $\psi_t$ is a normalized state vector.
For pure states, one can split $C$ into Hermitian and anti-Hermitian parts and re-write the diffusion term in~\cref{eq:sme_general_ito} as $\{C - \braket{A}_t, \rho_t\}_c = [[A, \rho_t] + B, \rho_t]$, where
\bal{\label{eq:B}
B \coloneqq (C - C^\dagger)/2
}
is the anti-Hermitian part of $C$.
By defining the state-dependent Hamiltonian $H_t \coloneqq i([A, \rho_t] + B)$, Eq.~\cref{eq:sme_general_ito} can be equivalently written as
\bal{
\label{eq:sme_ito}
d\rho_t &= \cL(\rho_t) dt -i[H_t, \rho_t]dW_t\,.
}
The form of~\cref{eq:sme_ito} is somewhat unconventional compared to the more common expression in Eq.~\cref{eq:sme_general_ito}, in that the noise term in~\cref{eq:sme_ito} appears as a Liouville contribution generated by the Hamiltonian $H_t$.
This reformulation is a direct consequence of the fact that the state remains pure at all times.
See App.~\ref{app:ito} for a derivation of~\cref{eq:sme_general_ito} and~\cref{eq:sme_ito} from the standard stochastic Schr\"odinger equation.

The continuous measurement process associated to Eqs.~\cref{eq:sme_general_ito} and~\cref{eq:sme_ito} is characterized by the output
\bal{
\label{eq:dyt}
dy_t = 2 \gamma \braket{A}_t dt + dW_t\,,
}
where $y_t$ is the accumulated measurement outcome~\cite{barchielliQuantumTrajectoriesMeasurements2009}.
Since $dW_t$ is of zero mean, upon averaging measurement instances over noise processes on a fixed state $\rho_t$, the noise-averaged measurement output uncovers the observable's expected value, $\overline{y_t} \propto  \braket{A}_t$.

Consider the pure-measurement case where $B = 0$ and $H_0 =0$.
The monitoring process is then characterized by the Hermitian observable $A$ and the rate $\gamma$.
We refer to this as the pure-measurement regime as, in such a case, subspaces corresponding to the eigen-projectors $\Pi_j$ of $A$ are fixed points of the dynamics.
Moreover, the system's state reaches such eigenspaces with probabilities given by the Born rule. The rate at which the full measurement process happens is determined by the rate $\gamma$ which is, in turn, dictated by interaction with the measurement apparatus.
See, e.g., Refs.~\cite{murch2008observation, murch2013observing, PhysRevX.6.011002} for experimental demonstrations of continuous measurement processes as described above.
Next, we relate the stochastic measurement dynamics described above to purely Hamiltonian processes.

%================================================================================

\vspace{7pt}
\section{Hamiltonian formulation of quantum measurements}
\label{sec:MeasurementHamiltonian}
Note that Eq.~\cref{eq:sme_ito} satisfies $\tr{d\rho_t} =d(\|\psi_t\|^2) = 0$, that is, the norm of the state is conserved throughout the entire evolution.
As a consequence, the system's state evolves unitarily, although stochastically, across the Hilbert space.
This suggests that a Hamiltonian interpretation of the monitored dynamics~\cref{eq:sme_ito} should be possible.
% Hamiltonian formulations that reproduce the paths traced by a set of connected pure states have been investigated in the context of the quantum brachistochrone~\cite{carliniTimeOptimalQuantumEvolution2006, huQuantumStateDriving2023} and counterdiabatic driving~\cite{alipour2020shortcuts}.
% Ref.~\cite{huDescribingWaveFunction2023} generalizes~\cite{huQuantumStateDriving2023} to continuous quantum measurements, and Ref.~\cite{hu2023probabilistic} formulates arbitrary open-system dynamics as a probabilistic unitary evolution.
However, such formulation is not feasible within the It\^o formulation of the dynamics; non-Hermitian corrections to the dynamics prevent it from being expressed as a quantum Liouville equation of the form
\bal{\label{eq:liouville}
d\rho_t = -i[dH_t, \rho_t]\,,
}
for some stochastic Hermitian generator process $dH_t$.

The situation changes in the Stratonovich picture, where the rules of stochastic calculus take the same form as those of standard calculus.
Although computations are generally less straightforward than in the Itô picture, the Stratonovich picture offers descriptions of stochastic processes that align more naturally with physical intuition~\cite{jacobsStochasticProcessesPhysicists, kloeden2012numerical}.
In particular, within the Stratonovich formulation of~\cref{eq:sme_ito}, we derive explicit expressions for the stochastic Hamiltonian governing the monitored dynamics in terms of the measurement outcomes.
Hamiltonian formulations that reproduce the paths traced by a set of connected pure states have been investigated in the deterministic case in the context of the quantum brachistochrone problem~\cite{carliniTimeOptimalQuantumEvolution2006, huQuantumStateDriving2023} and counterdiabatic driving~\cite{alipour2020shortcuts}.
For stochastic dynamics, Ref.~\cite{huDescribingWaveFunction2023} generalizes~\cite{huQuantumStateDriving2023} to provide a recipe for constructing retracing Hamiltonians for quantum measurement processes, and Ref.~\cite{hu2023probabilistic} formulates arbitrary open-system dynamics as probabilistic unitary evolutions.
More recently, Ref.~\cite{garcia-pintosReshapingQuantumArrow2025} derives an explicit Hamiltonian, for the Hermitian case $C=C^\dagger$, that retraces the monitored dynamics and uses it to define feedback protocols for steering the quantum arrow of time.
Our result generalizes that of~\cite{garcia-pintosReshapingQuantumArrow2025} to the case of arbitrary jump operators and differs from prior work in that the Hamiltonian is given explicitly in terms of jump operators and measurement outcomes.
Moreover, the dynamics is represented in terms of single- and double-bracket commutators in the state, which offers a new perspective between open quantum dynamics and double-bracket flows~\cite{Brockett1991}.

In what follows, we focus on processes whose dynamics are written in the Stratonovich form.
See App.~\ref{app:ito-strato} for a brief recap on the Stratonovich picture and It\^o-Stratonovich conversion formula.

We illustrate a straightforward connection between double brackets and quantum dynamics by noting that any stochastic dynamics that preserves a state's purity satisfies
\bal{\label{eq:db_master}
d\rho_t = [[d\rho_t, \rho_t], \rho_t]\,.
}
This follows directly by expanding the commutators and using $\rho_td\rho_t\rho_t = 0$, which itself is obtained by differentiating both sides of $\rho_t^2 = \rho_t$~\cite{teufelAdiabaticPerturbationTheory2003}.
It is straightforward to identify relation~\cref{eq:db_master} as a quantum Liouville equation of the form~\cref{eq:liouville}, with Hamiltonian generator given by $dH_t \coloneqq i[d\rho_t, \rho_t]$.
Note that the same reasoning holds true in the deterministic case.
Moreover, if $\rho_t$ is smooth, we have $\dot \rho_t = -i[H_t, \rho_t]$ with Hamiltonian $H_t = i[\dot \rho_t, \rho_t]$.

Equations~\cref{eq:db_master,eq:liouville} give a recipe for building the Hamiltonian generator of a predefined quantum trajectory $\rho_{0:T}$, where $T$ is the final time.
This procedure is the Liouville-space analog, for stochastic processes, of constructing Schrödinger equations that reproduce trajectories in Hilbert space~\cite{carliniTimeOptimalQuantumEvolution2006, huQuantumStateDriving2023}.
Conversely, a predefined Hamiltonian generator $dH_t$ yields the corresponding pure-state dynamics; for instance, setting $dH_t = H_0 dt$ gives the well-known quantum Liouville equation $\dot \rho_t = -i[H_0, \rho_t]$.

Observe that a generator process like $dH_t = H_0 dt$ yields dynamics in a single-bracket commutator form.
An immediate question is whether higher-order bracket contributions can be constructed that still preserve the purity of the state.
Under the assumption of purity preservation $\rho_t^2 =\rho_t$, it follows that single- and double-bracket terms constitute the most general possible contributions.
To see this, consider the following decomposition
\bal{\label{eq:multi_bracket_1}
d\rho = \cC_\rho (d\Lambda_1) + \cC_\rho^2 (d\Lambda_2) + \cC_\rho^3 (d\Lambda_3) + \ldots
}
where, for the moment, we omit the time subscript.
The super-operator $\cC_\rho$ is defined as $\cC_\rho(\bullet) := [\bullet,\,\rho]$.
The quantities $d\Lambda_i\,(i=1, \ldots)$ are arbitrary (not necessarily Hermitian) generator processes.
For pure states, the super-operator $\cC_\rho$ satisfies the periodic property $\cC_\rho^{2k+1} = \cC_\rho\, (k = 0, 1, \ldots)$.
This further implies that $\cC_\rho^{2k} = \cC_\rho^2\,(k=1,\, 2,\ldots)$.
See App.~\ref{app:periodic_property} for a proof.
We note that the periodic property of $\cC_\rho$ was exploited in various geometric and adiabatic settings (see, e.g., Ref.~\cite{teufelAdiabaticPerturbationTheory2003} and related projector-based constructions), and appears in similar form in recent double-bracket algorithms for signal processing~\cite{suzuki2025double}.
Using the periodic property we write~\cref{eq:multi_bracket_1} equivalently as
\baln{
d\rho = \cC_\rho (d\Lambda_1 + d\Lambda_3 + \ldots) + \cC_\rho^2 (d\Lambda_2 + d\Lambda_4 + \ldots)\,.
}
We conclude that the most general multi-bracket expansion for pure dynamics is equivalent to a single- and double-bracket decomposition of the form $d\rho_t = -i [dH_t^\sb, \rho_t] + [[dH_t^\db, \rho_t], \rho_t]$, where $dH_t^\sb$ and $dH_t^\db$ are some Hermitian generator processes.

Next, we present a result with explicit expressions for the generators $dH_t^\sb$ and $dH_t^\db$, when the dynamics corresponds to monitored open quantum systems governed by~\cref{eq:sme_ito}.
These expressions are in terms of the Lindblad jump operator $C$ and the measurement outcomes $dy_t$.
We establish the following result.
\begin{theorem}\label{th:sme_strato}
    The monitored dynamics~\cref{eq:sme_ito} in Stratonovich form is given by the stochastic quantum Liouville equation
\bal{\label{eq:sme_strato}
    d\rho_t = -i [dH_t^\sb, \rho_t] + [[dH_t^\db, \rho_t], \rho_t],
}
where the single-bracket and double-bracket Hamiltonian generator processes are, respectively,
\bal{\label{eq:Hsb}
    dH_t^\sb \coloneqq H_0 dt -i\frac{\gamma}{2} \{A, B\} dt + i B dy_t
}
and
\bal{\label{eq:Hdb}
dH^\db_t \coloneqq -\gamma A^2 dt -\frac{\gamma}{2}[A, B] dt + A dy_t\,,
}
where $A$ and $B$ are defined in Eqs.~\cref{eq:A} and~\cref{eq:B}.
\end{theorem}
Equation~\cref{eq:sme_strato} should be interpreted according to the midpoint evaluation rule characteristic of the Stratonovich picture~\cite{oksendal2003stochastic, kloeden2012numerical} (see also App.~\ref{app:ito-strato} for a recap of definitions).
We leave the proof of Theorem~\ref{th:sme_strato} and its extension to multiple noise channels to the Appendix~\ref{app:theorem_1}.

We emphasize the physical interpretation of Eq.~\cref{eq:sme_strato}, which recasts open quantum dynamics into a stochastic Hamiltonian evolution of the form given in Eq.~\cref{eq:liouville}, generated by the Hermitian differential process
\bal{\label{eq:dH}
dH_t = dH_t^\sb + i[dH_t^\db, \rho_t]\,.
}
This representation remains valid for an arbitrary Lindblad jump operator $C$.
On the other hand, the Hamiltonian process $H_t^\sb = \int_0^t dH_\tau^\sb$ can be interpreted as a modification of the system Hamiltonian $H_0$ by the noise back-action due to the system-environment interaction conditioned on the measurement process.
Conversely, the Hamiltonian process $H_t^\db = \int_0^t dH_\tau^\db$ encodes the measurement back-action resulting from the measurement process.

Note that both processes $H_t^\sb$ and $H_t^\db$ depend on the state $\rho_t$ only via the measurement process $dy_t$.
However, their experimental realization differs for the two processes.
In the single-bracket case, $dH_t^\text{SB}$ itself acts as the Hamiltonian generator.
In the double-bracket case, by contrast, the effective Hamiltonian generator is $i[dH_t^\db, \rho_t]$.
The first contribution admits a straightforward implementation using standard techniques, whereas the second is explicitly state dependent and therefore requires (fast) parallel state estimation; this is in principle achievable using the conditioned dynamics of the state, but comes with practical limitations such as measurement delays and inefficiencies, model inaccuracies, and finite computational precision.
We return to this point later in Section~\ref{sec:Feedback}.

We note that, in the Hermitian case $C = C^\dagger$, Eq.~\cref{eq:sme_strato} reduces to the monitored dynamics derived in Ref.~\cite{garcia-pintosReshapingQuantumArrow2025}.
See discussion around~\cref{eq:sme_hermitian_case} in App.~\ref{app:strato} for the explicit correspondence.

In general, quantum Liouville equations are written in Lax form~\cite{blochCompletelyIntegrableGradient1992}, which are isospectral flows preserving the eigenvalues of the density operator.
On the other hand, it is known that Itô diffusions of the form~\cref{eq:sme_ito} tend to asymptotically purify initially mixed states~\cite{barchielliQuantumTrajectoriesMeasurements2009}.
This implies that the spectrum of mixed density operators is not preserved during the evolution, and a Hamiltonian formulation of the dynamics is no longer applicable.

We remark that, while the monitored state $\psi_t$ traces a unitary path across the Hilbert space, it does not mean that the corresponding stochastic propagator [see~\cref{eq:stoch_prop} in the Appendices] is a valid unitary operator for arbitrary states, since its definition depends explicitly on the monitored state.

\begin{figure}
    \centering
    \includegraphics[width=0.8\columnwidth]{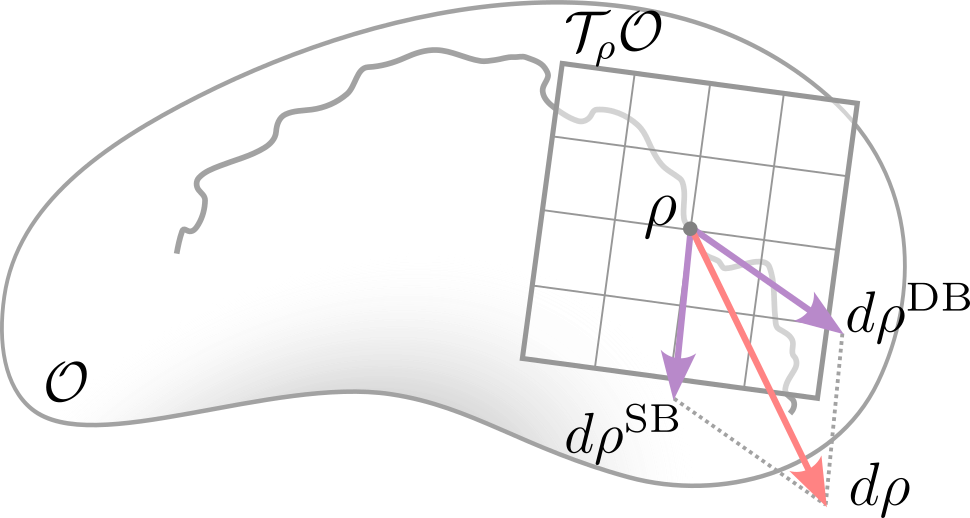}
    \caption{
        Illustration of the Hamiltonian evolution given by~\cref{eq:sme_strato}.
        The evolution takes place on the unitary orbit $\cO$ of the density operator.
        In the tangent space $\cT_\rho \cO$ at point $\rho$, the increment $d\rho$ is composed of two (not necessarily orthogonal) contributions, $d\rho^\sb = -i[dH_t^\sb, \rho_t]$ and $d\rho^\db = [[dH_t^\db, \rho_t], \rho_t]$, corresponding to the single-bracket and double-bracket Hamiltonians, respectively.
        }
    \label{fig:ham_flow}
\end{figure}

Equation~\cref{eq:sme_strato} comes equipped with a direct geometrical interpretation.
The evolution takes place in the so-called unitary orbit of the state $\rho$, defined by $\cO_\rho = \{U \rho U^\dagger | U\, \text{unitary}\}$.
At any given time, the increment $d\rho$ takes place in the tangent space of the orbit at point $\rho$, $\cT_\rho \cO_\rho$~\cite{knappLieGroupsIntroduction1996,nakaharaGeometryTopologyPhysics2003, rossmannLieGroupsIntroduction2006}, and is composed of single- and double-bracket contributions, $d\rho^\sb$ and $d\rho^\db$, respectively, each of which is a commutator with the state.
Thus, the Hamiltonian processes are the generators of dynamical flows in the orbit of the state.
We illustrate this geometrical picture in Fig.~\ref{fig:ham_flow}.
We return to this picture in the next section in the context of gradient flows.

%================================================================================

\vspace{7pt}
\section{Connection with double-bracket gradient flows}
\label{sec:DoubleBracket}

\subsection{Background}

Gradient flows represent a powerful approach to constrained optimization problems~\cite{HelmkeMoore1994, blochHamiltonianGradientFlows1994}.
Generically defined as evolutions on a constrained manifold, gradient flows follow paths of steepest descent (or ascent) of a given quality function defined on that manifold.
In quantum applications, gradient flows play a crucial role in enabling advanced methods for quantum control~\cite{dirr2008lie, schulte-herbruggenGradientFlowsOptimization2010} and quantum machine learning~\cite{wiersemaOptimizingQuantumCircuits2023, wiersemaHereComesSUN2024}.

On the other hand, double-bracket (DB) flows have, in recent years, gained increasing attention within the quantum computing community for the design of efficient quantum algorithms~\cite{schulte-herbruggenGradientFlowsOptimization2010, xiaoyueStrategiesOptimizingDoublebracket2024, gluzaDoublebracketQuantumAlgorithms2024,gluzaDoublebracketQuantumAlgorithms2025}.
DB flows are commonly introduced as equations of the form $dX/dt = [[N, X(t)], X(t)]$ where $N$ is a diagonal matrix with unique entries.
These systems exhibit the remarkable property that the unique stable fixed point $X_\infty = \lim_{t \goto \infty}X(t)$ of the dynamics is also a diagonal matrix with elements following the same sorting as those of $N$.
In other words, if $\text{diag}(N)$ is sorted in increasing order, $\text{diag}(X_\infty)$ is in increasing order as well~\cite{Brockett1991}.
Double-bracket flows have also been studied in the quantum physics literature in the setting of continuous renormalization-group methods for Hamiltonian diagonalization~\cite{glazek1993renormalization, wegner1994flow, hornedal2023geometric}.

In geometrical terms, DB flow equations can be formulated as gradient flows on the adjoint orbit of a Lie group~\cite{blochCompletelyIntegrableGradient1992, tam2004gradient, schulte-herbruggenGradientFlowsOptimization2010}.
This connection was recently exploited in Refs.~\cite{gluzaDoublebracketQuantumAlgorithms2025, mcmahon2025equating} to frame imaginary-time evolutions~\cite{goldberg1967integration} as DB flows for ground state preparation, enabling coherent implementations on quantum hardware.
In this context, DB flows provide the crucial theoretical bridge between the fundamentally non-unitary operation defined by the imaginary-time evolution and a circuit implementation in terms of unitary operations.

More broadly, it is noteworthy that connections between DB gradient flows and quantum-mechanical evolutions are often encountered in the literature through imaginary-time evolution rather than through real-time quantum dynamics.
Ref.~\cite{mittnenzweig2017entropic} shows that detailed-balance Lindblad generators admit an entropic gradient-flow formulation and can be represented through a lifted double-commutator structure. Our result is different in nature: in the Stratonovich picture, the double-bracket structure acts directly on the evolving state, with each such term expressed as a Riemannian gradient on the unitary orbit of $\rho_t$.
Thus, it is not surprising that this connection can be taken a step further by framing the quantum Liouville equation~\cref{eq:sme_strato} in the language of gradient flows.
Building on Theorem~\ref{th:sme_strato}, we instead write each double-bracket term in~\cref{eq:sme_strato} as a Riemannian gradient on the unitary orbit of $\rho_t$.

\subsection{Results}

The Hilbert-Schmidt (HS) metric on the unitary group $U(N)$ is given by the inner product $\langle X,Y \rangle_{\textnormal{HS}} \coloneqq \tr{X^\dag Y}$, where $X,\, Y$ belong to the Lie algebra of $U(N)$.
The HS metric restricted to the unitary orbit~\cite{schulte-herbruggenGradientFlowsOptimization2010} defines a Riemannian metric $\braket{\bullet,\, \bullet}_\cO$ such that $\braket{X,\,Y}_{\textnormal{HS}} = \braket{X',\, Y'}_\cO$, where $X' = [X, \rho]$ and $Y' = [Y, \rho]$ are tangent vectors on the state's orbit.

To motivate the subsequent discussions, consider Eq.~\cref{eq:db_master} in the deterministic case, that is, $\dot \rho_t = [[\dot\rho_t, \rho_t],  \rho_t]$.
By defining the scalar potential $F_t(\sigma) = \tr{\dot \rho_t \sigma}$ with $\sigma$ in the state's orbit, one can prove that the dynamics can be written as the gradient flow $\dot \rho_t = \nabla F_t(\rho_t)$, where $\nabla F_t(\rho_t)$ is the Riemannian gradient induced by the HS metric on the state's orbit.

A more general result turns out to be true for stochastic processes.
\begin{prop}\label{prop:stoch_flow}
A DB equation of the form $d\rho_t = [[dO_t, \rho_t], \rho_t]$, with $dO_t$ some Hermitian generator process, admits the gradient flow interpretation $d\rho_t = \nabla F_t(\rho_t) dt$, with potential given by $F_t(\sigma) = \tr{\dot O_t \sigma}$ and $\dot O_t = dO_t/dt$.
\end{prop}
This general result is, to the best of our knowledge, new, and is a direct consequence of the double-bracket form of the dynamics.
We refer the reader to App.~\ref{app:gradient_flow} for the full derivation.

Note that the deterministic case $\dot \rho_t = [[\dot \rho_t, \rho_t], \rho_t]$ is a particular instance of Prop.~\ref{prop:stoch_flow} in the absence of noise and when $O_t \equiv \rho_t$.

Proposition~\ref{prop:stoch_flow} is to be interpreted in the sense of stochastic differential equations.
A general Hermitian generator process $dO_t$ is of the form $dO_t = M_t dt + \Sigma_t dW_t$, for some matrix-valued processes, $M_t$ and $\Sigma_t$.
As a consequence, $F_t(\sigma)dt = \tr{M_t \sigma} dt + \tr{\Sigma_t \sigma} dW_t$, and we write the dynamics as $d\rho_t = \nabla F_t^{(1)}(\rho_t) dt + \nabla F_t^{(2)}(\rho_t) dW_t$ with $F_t^{(1)}(\sigma) = \tr{M_t \sigma}$ and $F_t^{(2)}(\sigma) = \tr{\Sigma_t \sigma}$.

A direct application of Prop.~\ref{prop:stoch_flow} to the Hamiltonian dynamics~\cref{eq:sme_strato} leads to the following interpretation of measurement processes as gradient flows.
\begin{theorem}\label{th:grad_flow}
    The monitored dynamics~\cref{eq:sme_strato} can be written as a gradient flow on the unitary orbit as
    \bal{\label{eq:flow}
    d\rho_t = \nabla F_t (\rho_t) dt,
    }
    where the potential $F_t$ consists of the sum of two parts,
    \bal{\label{eq:potentialSB}
    F_t^\sb(\sigma) = \tr{-i[\dot H_t^\sb, \rho_t] \sigma}
    }
     and
     \bal{
     \label{eq:potentialDB}
     F_t^\db(\sigma) = \tr{ \dot H_t^\db \sigma}\,.
     }
    $\dot H_t^\sb \equiv dH_t^\sb/dt$ and $\dot H_t^\db \equiv dH_t^\db/dt$ are the instantaneous single- and double-bracket Hamiltonians, respectively, and $\sigma$ belongs to the state's orbit.
\end{theorem}
Consult App.~\ref{app:gradient_flow} for the proof.

\begin{figure}
    \centering
    \includegraphics[width=0.5\columnwidth]{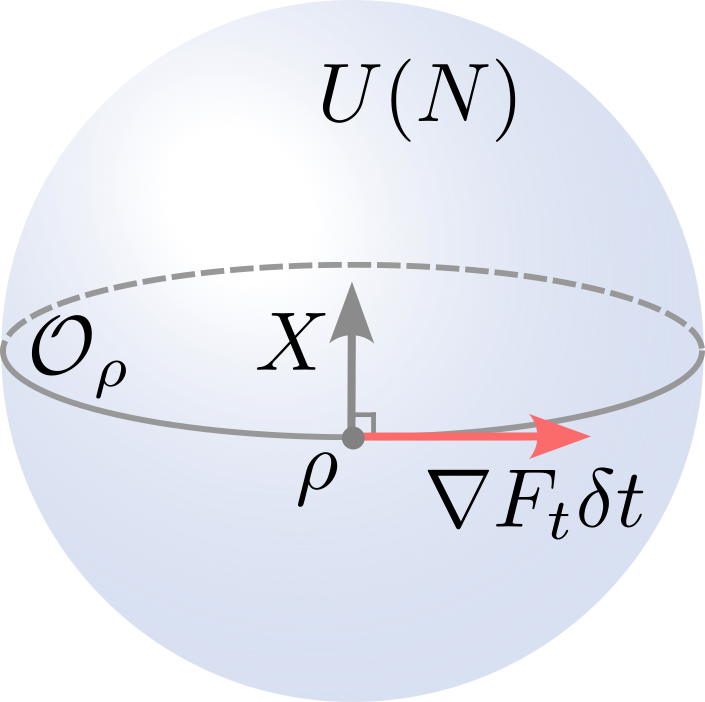}
    \caption{
    Schematic representation of the gradient flow formulation in Theorem~\ref{th:grad_flow}.
    We illustrate the embedding of the state's orbit $\cO_\rho$ (circumference), in the unitary group $U(N)$ (sphere).
    The tangent vectors to the orbit correspond to vectors in the Lie algebra that are orthogonal to the stabilizer algebra of $\rho$, which is generated by vectors $X$ that commute with $\rho$.
    In particular, the gradient flow generates displacements $\delta \rho_t = \nabla F_t \delta t$ tangent to the orbit.
    }
    \label{fig:embedding}
\end{figure}

In Fig.~\ref{fig:embedding}, we illustrate the flow formulation given by Theorem~\ref{th:grad_flow}.
To facilitate an intuitive picture of the underlying geometry, we can think of the state's orbit as an embedded submanifold of the unitary group, with the latter being represented by a sphere.
The state's orbit is represented by a circumference, whose tangent vectors are orthogonal to those vectors $X$ in the Lie algebra that commute with the state, i.e.,
the generators of the stabilizer group~\cite{nakaharaGeometryTopologyPhysics2003}.
These vectors represent directions in the unitary group that do not generate any dynamics, that is, unitary operations such that $U \rho_t U^\dagger = \rho_t$.

To gain a deeper physical understanding of Eq.~\cref{eq:flow}, we write the flow equation in the equivalent form
    \bal{\label{eq:equiv_flow}
        d\rho_t = \nabla F_t^\sb(\rho_t) dt
        -\gamma \nabla V_t(\rho_t) dt
        + \nabla K(\rho_t) dW_t\,,
    }
where we define the potentials
\baln{
V_t(\sigma) \coloneqq \tr{\Delta A_t^2 \sigma} + \frac{1}{2}\tr{[A, B] \sigma}
}
with $\Delta A_t \coloneqq A - \braket{A}_t$, and
\baln{
K(\sigma) \coloneqq \tr{A \sigma}\,.
}
The equivalence between both flows is due to the $U(1)$ gauge freedom reflected in the identity $\cC_\rho(dO_t + d\alpha_t) = \cC_\rho(dO_t)$ for an arbitrary scalar process $\alpha_t$, since adding $d\alpha_t$ would introduce an unobservable phase in the physical state.

The potentials $V_t$ and $K$ are quantum expectations of various observables defined in terms of $A$ and $B$.
We write $V_t$ with a time subscript to emphasize that its definition depends on $\rho_t$, with domain in the unitary orbit of $\rho_t$.
The value $V_t(\rho_t) = \tr{\Delta A_t^2 \rho_t} + \frac{1}{2}\tr{[A, B] \rho_t}$ has two contributions with direct physical interpretation.
The first term corresponds to the variance of the observable $A$.
The second term can be interpreted as the lower bound of the Robertson inequality~\cite{robertsonUncertaintyPrinciple1929} between $A$ and $B$, that is,
$-\braket{\Delta A_t^2} \braket{\Delta B_t^2} \geq |\braket{[A, B]}|^2/4$, where $\Delta B_t \coloneqq B - \braket{B}_t$ (the minus sign is due to the anti-hermiticity of $B$).

In the diffusion term, the potential $K(\rho_t)$ represents the quantum expectation of the observable $A$.
At an arbitrary point $\rho$, the gradients of these functions are vectors tangent to the state's orbit.
Because~\cref{eq:equiv_flow} is stochastic, one can interpret the gradient $\nabla V_t(\rho_t)$ as inducing non-zero drifts in the orbit in directions where the variation in the quantum expectations are the steepest.
Conversely, the gradient corresponding to the diffusion term does not have a preferred direction due to the undefined sign of the Wiener increment $dW_t$.

On the other hand, the single-bracket potential  $F^\text{SB}_t$ in Eq.~\cref{eq:potentialSB} captures the contribution of the single-bracket Hamiltonian $H^\text{SB}$ to the gradient flow, and thus quantifies how the coherent drive reshapes the measurement-induced landscape on the unitary orbit.
Note that, when $A=0$, there is no measurement and no double-bracket term.
Thus, the total Hamiltonian process reduces only to the single-bracket contribution $dH^\text{SB}_t = H_0 dt + iB dW_t$, where the stochasticity arises from background noise generated by the system-environment interaction.

Let us next focus on the pure-measurement case where $B=0$ and $H_0 = 0$.
The single-bracket term in~\cref{eq:equiv_flow} vanishes, and the dynamics, expressed in terms of DBs, are described by
\bal{\label{eq:meas_grad_flow}
    d\rho_t = -\gamma [[\Delta A_t^2, \rho_t], \rho_t] dt + [[A, \rho_t], \rho_t] dW_t\,.
}
The stationary points of~\cref{eq:meas_grad_flow} coincide with the eigenstates of $A$, as they should for pure measurement processes.
One can show by computing the evolution equation of the overlap between the state and any of the eigenspaces of $A$ that these subspaces are attractors of the dynamics, and therefore stable.
Moreover, when the state is close to an eigenspace of $A$, the variance in the overlap tends to vanish, making it harder for the system to escape from the vicinity of the attractor subspace.

\begin{figure}
    \centering
    \includegraphics[width=0.8\columnwidth]{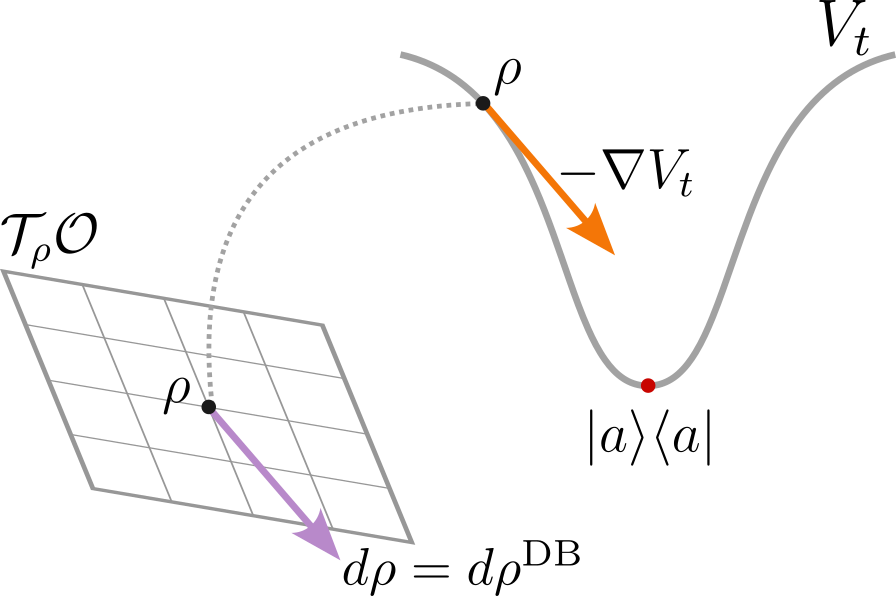}
    \caption{
        Illustration of the pure-measurement case $C = A$ and $H_0 = 0$.
        The increment $d\rho$ is composed only of the double-bracket term $d\rho^\db$, with drift dictated by the Riemannian gradient of the the variance $V_t(\rho_t) = \text{Tr}(\Delta A_t^2 \rho_t)$.
        The variance vanishes when $\rho_t$ corresponds to an eigenstate $\ket{a}$ of $A$.
        }
    \label{fig:basin}
\end{figure}

Alternatively, these observations can be drawn from the gradient interpretation of~\cref{eq:meas_grad_flow}.
The drift term, which is equivalent to $-\gamma \nabla V_t(\rho_t)dt$, steers the state in directions that minimize the variance $V_t(\rho_t)$ associated to the operator $A$.
This can be seen more clearly by computing the Stratonovich differential of $V(t) \coloneq V_t(\rho_t) = \tr{\Delta A_t^2 \rho_t}$, which gives
\bal{\label{eq:strato_variance}
 dV(t) = -\gamma \|\nabla V_t(\rho_t)\|_\cO^2 dt + \braket{\nabla V_t(\rho_t), \nabla K(\rho_t)}_\cO dW_t\,,
}
where we used the compatibility condition of the metric (see App.~\ref{app:math_back} for definitions) and the pure-measurement dynamics $d\rho_t = -\gamma \nabla V_t(\rho_t) dt + \nabla K(\rho_t) dW_t$.
See App.~\ref{app:variance} for the derivation.
Equation~\cref{eq:strato_variance} says that the variance of the measurement operator tends to decrease along the solution trajectory with rate proportional to the squared norm of its gradient on the orbit.
The competing term $\braket{\nabla V_t(\rho_t), \nabla K(\rho_t)}_\cO dW_t$ introduces stochasticity in the dynamics, preventing full localization of all trajectories.
However, at hitting time, i.e. when the system enters an eigenspace of the measuring operator, the stochastic term vanishes exactly, since $\nabla V_t(\rho_t) = 0$.
Since the minima of the potential $V_t$ correspond to the eigenspaces of $A$, the long-time dynamics will localize the state in one of these eigenspaces, and the measurement apparatus will retrieve the corresponding eigenvalue almost surely.
See Fig.~\ref{fig:basin} for a schematic illustration of this idea.
One must note the resemblances with Lyapunov stability~\cite{kloeden2012numerical}, where the potential $V_t$ can be related to the Lyapunov function of the dynamics.

We note that the monitored dynamics tends to minimize the variance of the measurement operator is known from previous work; see, for instance, Refs.~\cite{AdlerBrodyBrunHughston2001, BrodyHughston2002JMP, BrodyHughston2006JPA, adlerCollapseModelsNonwhite2007}.
The novelty of Eq.~\cref{eq:meas_grad_flow}, or its gradient flow equivalent, is that it gives a streamlined interpretation of the monitored dynamics as a continuous collapse process, where the asymptotic fate of the state is a stationary density operator with the correct Born probabilities~\cite{Belavkin1989PLA,BarchielliBelavkin1991JPA, jacobsStraightforwardIntroductionContinuous2006}.

When the system Hamiltonian $H_0$ is non-zero, the asymptotic collapse analysis remains valid under the assumption of quantum non-demolition measurements, i.e. when $H_0$ commutes with the measurement operator $A$~\cite{jacobsStraightforwardIntroductionContinuous2006, bauerRepeatedQuantumNondemolition2013}. When $H_0$ and $A$ do not commute, localization is still possible in non-trivial decoherence-free subspaces~\cite{schmolkeAsymptoticFateContinuously2025}.

It is straightforward to generalize Eq.~\cref{eq:meas_grad_flow} to multiple measurement operators $A_i$ as a linear combination of DB flows contributions, one per noise channel.
In this case, the different flows compete to reduce the variance of each measurement, with localization being possible in the common eigenspaces associated with all measurement operators.

In the most general case where $H_0 \neq 0$, $A\neq 0$, and $B \neq 0$, Eq.~\cref{eq:sme_strato} involves various competing terms that prevent collapse unless special conditions are met.
For instance, when $A$ and $B$ commute.
Assuming the existence of non-trivial common eigenspaces between $H_0$ and $A$, one can show that these subspaces are attractors of the dynamics by studying the behavior of the overlap of the system's state with any of the subspaces.
In particular, if the state enters a common eigenspace corresponding to an eigenvalue $a$, the dynamics becomes the quantum Liouville equation~\cref{eq:liouville} with Hamiltonian $dH_t \coloneqq iB_a (-\gamma a dt + dy_t)$, where $B_a = \Pi_a B \Pi_a$ is the orthogonal projection of $B$ onto the eigenspace of $A$ with eigen-projector $\Pi_a$.
Thus, the effect of the anti-Hermitian operator $B$ is to unitarily scramble the state within the selected eigenspace.
Setting the hitting time in the selected eigenspace to $t=0$, with corresponding state $\rho_0$, we can integrate the dynamics and write the solution as $\rho_t = U(t)\rho_0U(t)^\dagger$, with the propagator $U(t) = \exp(B_a (-\gamma a t + y_t)) $ and the accumulated measurement outcome $y_t = \int_0^t dy_s$.

For general non-commuting $A,\, B$, the variance of $A$ is lower bounded by the Robertson inequality $-\braket{\Delta A_t^2} \braket{\Delta B_t^2} \geq |\braket{[A, B]}|^2/4$, which might prevent further localization.
Nevertheless, this lower bound is exactly one of the quantities that the dynamics tends to minimize, together with the variance of the observable $A$.
As a consequence, a synergy emerges between the two DB terms such that the variance in $A$ gets minimized as much as the minimization of the Robertson bound allows.
We already described a related phenomenon in the previous paragraph assuming the stronger condition of commuting $[A, B]= 0$.
In that case, the dynamics is free to localize the state in any of the common eigenspaces of $H_0$ and $A$.
In the non-commuting case, the dynamics will still promote (although incomplete) localization by simultaneously minimizing the Robertson bound and the variance in $A$.
This inevitably resembles the case of coherent states for continuous measurement operators.
For the harmonic oscillator with jump operator $C= X + iP$ where $X,\, P$ are the position and momentum operators, respectively, the monitored dynamics asymptotically converges to coherent states~\cite{jacobsStraightforwardIntroductionContinuous2006}.
In this case, the Robertson inequality with $A=X$ and $B=iP$ becomes increasingly saturated, and the state becomes increasingly coherent, simultaneously reducing the uncertainty in $A$ and $B$.

%================================================================================

\vspace{7pt}
\section{Applications to feedback control}
\label{sec:Feedback}

\subsection{Feedback-induced DB flows}
\label{sec:feedback_induced_DB}

We present a notable case where the stochastic monitored dynamics~\cref{eq:sme_strato} can be turned into a deterministic DB flow equation after incorporating feedback.

Consider the Hermitian case where $B = 0$.
Assume that the measurement apparatus couples to the system's energy operator, i.e., $A = H_0$.
The measurement process is characterized by a measurement record $dy_t = 2\gamma \braket{H_0}_t dt + dW_t$.
After each measurement at time $t$, we inject into Eq.~\cref{eq:sme_strato} the coherent feedback Hamiltonian 
\bal{\label{eq:fb1}
dH^\text{fb}_t = i[H_0, \rho_t] u_t dt\,,
}
where $u_t$ is a real number representing the control signal.
We design the feedback such that it precisely counteracts the measurement's influence on the system dynamics, achieved by setting $u_t dt = -dy_t$.
Under this feedback scheme, the system evolves according to the equation
\bal{\label{eq:sme_feedback}
    \frac{d\rho_t}{dt} = -i[H_0, \rho_t] - \gamma \left[ [H_0^2, \rho_t], \rho_t \right]\,.
}

The effect of counteracting the measurement process is to suppress the stochasticity in the dynamics, which subsequently leads to the deterministic equation~\cref{eq:sme_feedback}.
This scheme can be seen as a noise-canceling protocol.
In related work, Ref.~\cite{karmakarNoiseCancelingQuantumFeedback2025} uses coherent feedback to cancel the background noise due to the environment, and Ref.~\cite{garcia-pintosReshapingQuantumArrow2025} uses coherent feedback to stabilize the state of a system, thereby withdrawing energy from the monitoring process.
We use instead a feedback control to suppress the measurement back-action, which leads to a deterministic feedback-induced DB dynamics that, as we see below, can be used for ground state preparation.

Equation~\cref{eq:sme_feedback} comprises a deterministic isospectral evolution for which we can immediately apply known results from double-bracket flow theory.
Without loss of generality, consider the spectrum of $H_0$ to be $0 \leq E_0 < E_1 \leq \ldots \leq E_N$ with unique ground-state energy $E_0$.
We establish the following result.
\begin{theorem}\label{th:feedback}
    The unique stable fixed point of the feedback dynamics~\cref{eq:sme_feedback} is given by the ground state of $H_0$.
\end{theorem}
One can prove this in a number of ways.
One is transforming to the interaction picture, and invoking the results by Brockett~\cite{Brockett1991} and Helmke~\cite{HelmkeMoore1994} on asymptotic stability of DB flows.
Another way is analyzing the behavior of populations for the different energy levels and show that the ground state is the only stable asymptotic state; we show this in App.~\ref{app:simple_proof}.

An important feature of double-bracket flow evolutions is the exponential convergence to the stable solution~\cite{xiaoyueStrategiesOptimizingDoublebracket2024}.
In particular, the feedback protocol described above converges exponentially fast to the ground state of the system~\cite{HelmkeMoore1994} provided there is an initial non-zero overlap with the ground state~\cite{gluzaDoublebracketQuantumAlgorithms2024, gluzaDoublebracketQuantumAlgorithms2025}.

\begin{figure}
    \centering
    \includegraphics[width=0.9\columnwidth]{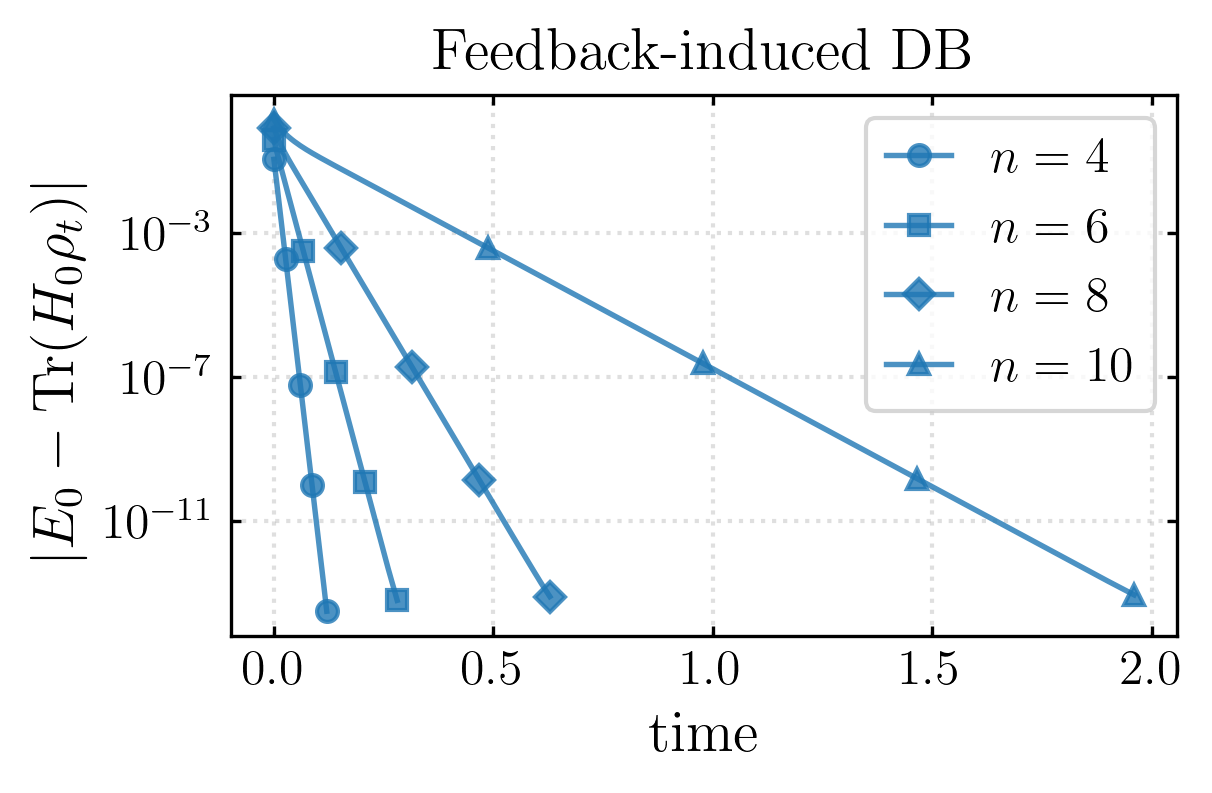}
    \caption{
    Energy errors for the 1-D Heisenberg model $H_0$~\cref{eq:heisenberg_model} and dynamics~\cref{eq:sme_feedback}.
    We monitor the error between the energy expectation $\tr{H_0 \rho_t}$ and the true ground state energy $E_0$ obtained from exact diagonalization.
    We do this for a total time $T=2$ and $300$ time steps.
    The damping rate is $\gamma = \num{4e-1}$.
    To ensure a positive spectrum, we shift the system Hamiltonian by adding a constant offset of $18$ for all qubit numbers.
    The $y$-axis is in logarithmic scale, showcasing the exponential convergence of the DB dynamics with errors reaching machine precision.
    }
    \label{fig:db_heisenberg}
\end{figure}

We illustrate the exponential convergence of the feedback-induced DB dynamics~\cref{eq:sme_feedback} for the 1-D antiferromagnetic Heisenberg model with Hamiltonian
\bal{\label{eq:heisenberg_model}
H_0 = \sum_{i=1}^{n-1} \lpar
\sigma^x_i \sigma^x_{i+1} + \sigma^y_i \sigma^y_{i+1} + \sigma^z_i \sigma^z_{i+1}
\rpar\,,
}
where $\sigma_i^x,\,\sigma_i^y,\, \sigma_i^z$ are Pauli strings.
This model exhibits a unique ground state for even numbers of qubits~\cite{LiebMattis1962, auerbach2012interacting}.
As initial condition we use (instead of a random state) a physically motivated~\cite{gluzaDoublebracketQuantumAlgorithms2025} tensor product state given by $\ket{\psi_0} \coloneqq 2^{-n/4} \lpar \ket{10} - \ket{01} \rpar^{\otimes n/2}$.
We evolve the system using the dynamics~\cref{eq:sme_feedback} over a certain time interval for different qubit numbers ($n=4, 6, 8, 10$), and monitor the error in computing the ground state energy $E_0$.
We show the results in Fig.~\ref{fig:db_heisenberg}.
We observe that, for the same damping rate value $\gamma$, the error decreases exponentially fast with time across different qubit numbers, reaching in all cases machine precision.

The feedback protocol described above leads to a concrete realization of dynamical ground-state preparation by harnessing properties of DB flows.
However, the fact that preparing the feedback Hamiltonian $dH^\text{fb}_t$ requires precise knowledge of the state can limit the applicability of the protocol, as its implementation would require real-time parallel state estimation.
This is feasible for low-dimensional systems, and has been the subject of active research~\cite{Doherty1999, Feedback2, Feedback3, Feedback6, Feedback7, Feedback8}.
By contrast, even if fast parallel estimation of the system's state is available, another challenge in scaling of the present feedback to large systems is the continuous measurement of the system’s energy, which generally requires coupling the meter to the full system Hamiltonian.
This is not impossible, however, as there are methods for performing weak measurements of the energy operator in many-body systems; see, e.g., ~\cite{yang2020quantum}.
We also emphasize that, even if these implementation challenges could be overcome, this would not imply an efficient worst-case algorithm for generic ground-state energy estimation, nor a resolution of the QMA-hard/QMA-complete local Hamiltonian problem~\cite{kitaevClassicalQuantumComputation2002, kempe2003local}.

To circumvent the requirement for parallel state estimation, an alternative approach is to forgo this step entirely and instead consider feedback Hamiltonians that, apart from knowledge of the measurement outcomes, remain completely agnostic about the state.

\vspace{7pt}
\subsection{State-agnostic feedback for state preparation}

Suppose we aim to prepare/stabilize an arbitrary pure state $Q = \ket{q}\!\bra{q}$. This time, we assume we only have access to the measurement outcomes $dy_t$, i.e., we don't know the system's state. We assume a setup with ideal measurement efficiency and no delay in the feedback loop. In what follows, we present a general blueprint for measurement-based feedback control that is agnostic of the state and produces dynamics that have as fixed point the state $Q$ we want to prepare or stabilize.

Consider a system evolving under the monitored dynamics~\cref{eq:sme_strato}.
Immediately after each measurement, we add to the dynamics two feedback Hamiltonians that depend, apart from system specifications, only on the current measurement outcome and the target state.
The first feedback is a counter-driving Hamiltonian process,
\baln{
dH_t^{\fb1} = -dH_t^\sb\,,
}
whose effect is to exactly cancel the action of the single-bracket Hamiltonian on the dynamics; see~\cref{eq:dH}.
The second feedback is
\baln{
dH_t^{\fb2} = -i[dH_t^\db,\,Q]\,,
}
i.e., the negative of the DB Hamiltonian in~\cref{eq:dH}, with the state $\rho$ replaced by $Q$.
Given these definitions, the effective dynamics is
\bal{\label{eq:sme_state_agnostic_fb}
d\rho_t =& - \gamma \lbra [A^2, \rho_t-Q], \rho_t \rbra dt
            -\frac{\gamma}{2} \lbra [[A, B], \rho_t-Q], \rho_t \rbra dt \nonumber \\
         &+ \lbra [A, \rho_t-Q], \rho_t \rbra dy_t\,.
}
The generalization of~\cref{eq:sme_state_agnostic_fb} to multiple noise channels is straightforward.

The state-agnostic feedback $dH^\text{fb}_t = dH^\text{fb1}_t + dH^\text{fb2}_t$ represents a coherent measurement-based feedback control that can, in principle, be implemented following well-established techniques in quantum control~\cite{wisemanQUANTUMMEASUREMENTCONTROL}.
A key distinction, however, is between the single-bracket and double-bracket contributions.
The single-bracket feedback $dH^\text{fb1} = -dH^\text{SB}_t$ is comparatively straightforward to implement, whereas the double-bracket feedback $dH^\text{fb2} = -i[dH_t^\db,\,Q]$ is more demanding, as it requires deliberate operator engineering to construct the relevant interaction terms, with the difficulty depending on the complexity of the target state $Q$.
For a single-qubit, for example, the DB feedback can be expressed as a sum of Pauli operators whose coefficients encode the measurement outcomes, the fixed state $Q$, and the jump operator/s.
More structured targets $Q$, especially when not available in closed form or defined only implicitly, as in the 1-D Heisenberg model of Sec.~\ref{sec:feedback_induced_DB}, require additional intermediate steps.
The advantage is that this pipeline naturally separates into offline and online stages: the operator engineering part can be done offline as a preprocessing step, while the online stage consists of processing the measurement outcomes in real time and feeding them back into the dynamics.

$Q$ is a fixed point of the dynamics, meaning  that whenever the state reaches the target, the dynamics immediately comes to a halt.
A natural question concerns the stability of $Q$ as a fixed-point solution.
The answer is problem-dependent, as we illustrate below.

Equation~\cref{eq:sme_state_agnostic_fb} can be interpreted in two different yet complementary ways.
In the first view, we are given an uncontrollable environment that can be probed to gather information about the system.
A stability analysis determines which regions of the Hilbert space are suitable for state preparation and stabilization.
In the second view, we are given a target state $Q$, and the task is to design suitable system-environment interactions such that the target state is a fixed point of the dynamics and can be certified as stable.
This perspective is closely related to dissipative reservoir engineering (also called bath or environment engineering), where one tailors the system–environment coupling so that the dissipative dynamics autonomously drives the system toward and stabilizes the desired target state~\cite{ticozzi2009analysis, barreiroOpenSystemQuantumSimulator2011, pan2016ground,
li2024autonomous, guoDesigningOpenQuantum2025, aydougan2025stabilizing, swain2025noiseassistedfeedbackcontrolopen}.
We illustrate the interplay of these two views with two examples.

\subsubsection{Example 1: the single-qubit}
\label{sec:single_qubit}

Assume a single-qubit system undergoing amplitude damping at a rate $\gamma_- > 0 $ and described by the jump operator $C = \sigma^- = (\sigma_x - i \sigma_y)/2$, where $\sigma_x,\, \sigma_y$ are Pauli operators.
The measuring apparatus is coupled to the Hermitian part of $C$ with measurement record given by $dy_t^- = \gamma_- \braket{\sigma_x}_t dt + dW_t^-$ and $(dW_t^-)^2 = \gamma_- dt$.
Assume we want to prepare a state $\ket{q}$ defined by Bloch angles $(\theta, \phi)$.
Ref.~\cite{wangFeedbackstabilizationArbitraryPure2001} analyzed a similar example with an alternative control scheme.
Consider the overlap process $dQ_\parallel(t) = \tr{Qd\rho_t}$ that quantifies the support of the state along the target.
A stability analysis of the Lyapunov exponent of the overlap process shows that the region of stability corresponds to states satisfying
$\cos(\theta) - 2 \sin^2(\theta) \cos^2(\phi) < 0$~(see App.~\ref{app:stability_analysis} for details).
A direct consequence of this relation is that the feedback protocol can stabilize arbitrary states in the lower hemisphere of the Bloch sphere, while most of the upper hemisphere remains unstable.

That most of the upper hemisphere is unstable is a natural consequence of the energy loss induced by the amplitude damping. To counteract this loss, we introduce a second feedback mechanism.
Assume that the system is coupled to an energy reservoir.
We model this by augmenting the previous setting with a jump operator $\sigma^+ = (\sigma_x + i \sigma _y)/2$, with damping rate $\gamma_+$.
The measurement record is given by $dy_t^+ = \gamma_+ \braket{\sigma_x}_t dt + dW_t^+$ with $(dW_t^+)^2 = \gamma_+ dt$.
Note that, although both jump operators lead to the same observable, we have two independent measurement streams representing different sources.
The feedback-induced dynamics for this configuration is
\bal{\label{eq:sme_single_qubit}
d\rho_t = \frac{\Delta \gamma}{4} \lbra [\sigma_z, \rho_t-Q], \rho_t \rbra dt
+ \frac{1}{2}\lbra [\sigma_x, \rho_t-Q], \rho_t \rbra dy_t\,,
}
where $\Delta \gamma \coloneqq \gamma_+ - \gamma_-$ and $dy_t = dy_t^+ + dy_t^-$.
By analyzing the stability of the fixed-point solution $Q$, we find that the stable region satisfies
\baln{
\Delta \gamma \cos(\theta) + 2 (\gamma_+ +\gamma_-) \sin^2(\theta) \cos^2(\phi) > 0\,.
}

For target states with angle $\phi = \pi/2$, stabilization in the upper hemisphere ($\theta < \pi/2$) requires $\Delta \gamma  > 0$; that is, energy injection must exceed dissipation due to damping.
For angles $\phi \neq \pi/2$, the condition $\Delta \gamma > 0$ is merely sufficient to achieve stability in the upper hemisphere.
The equatorial line $\theta =\pi/2$ also represents fixed points of the dynamics.
There, stability is achieved trivially for $\phi \neq \pi/2$.
This implies, as shown by simulations, that states close to (and including) the eigenstates of $\sigma_x$ are increasingly more stable than those near the eigenstates of $\sigma_y$.
Conversely, a sufficient condition for stabilization in the lower hemisphere is $\Delta \gamma < 0$; that is, energy renewal must not exceed the losses due to dissipation.

\begin{figure}
    \centering
    \includegraphics[width=0.9\columnwidth]{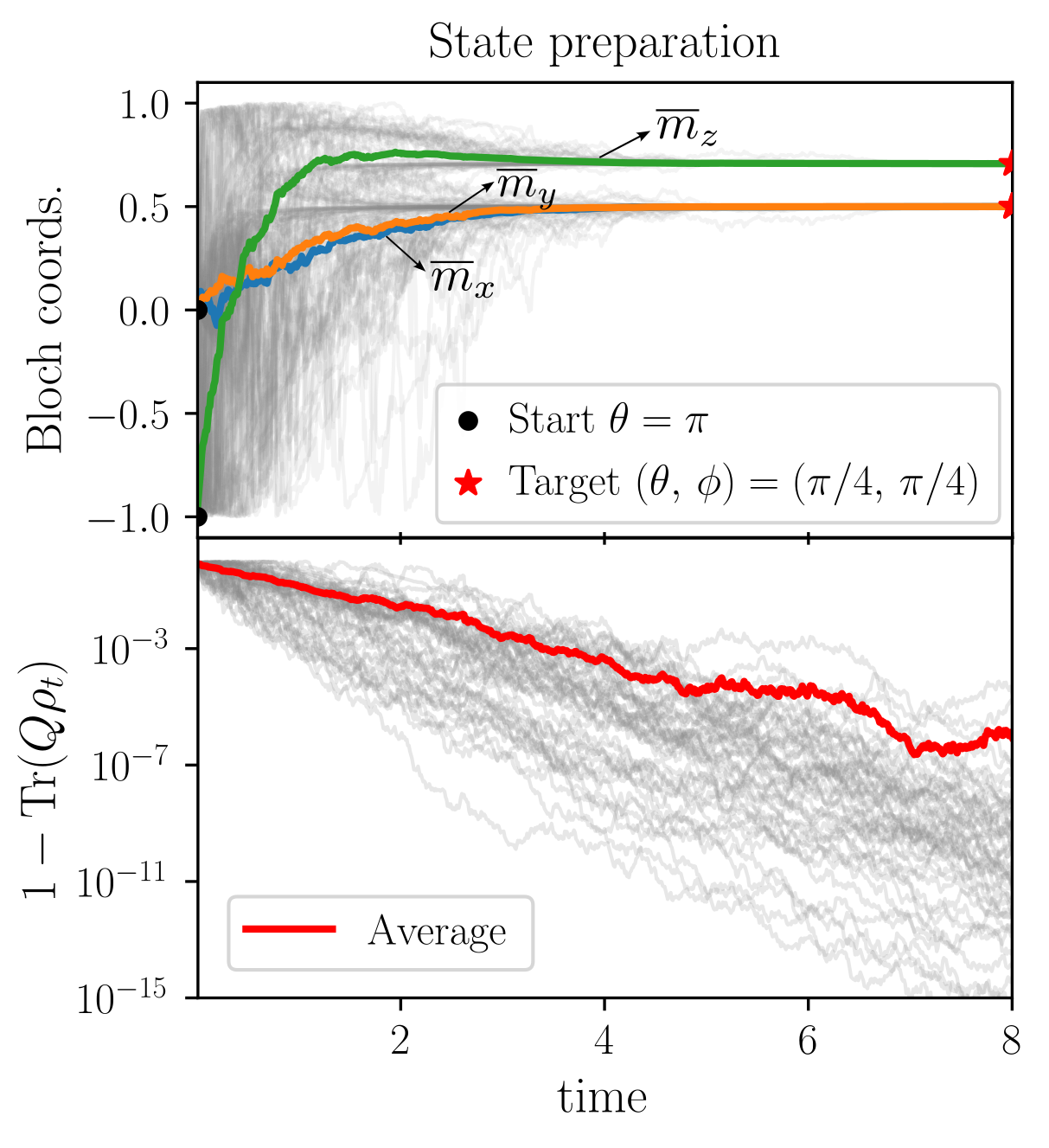}
    \caption{
    Measurement-based feedback state preparation and stabilization for the single-qubit with dynamics given by~\cref{eq:sme_single_qubit}.
    We display (top panel) the time evolution of the Bloch coordinates $m_i(t) = \tr{\sigma_i\rho_t}\,(i=x,\, y,\, z)$ of $50$ trajectories along their ensemble averages $\overline{m}_x,\,\overline{m}_y,\, \overline{m}_z$.
    The parameters are $\gamma_- = 1$ and $\gamma_+ = 4$, with final time $T=8$, for a total of $800$ time steps.
    All trajectories stabilize relatively fast around the target, with a final average infidelity (bottom panel) $1- \overline{\text{Tr}(Q \rho_T)} \sim 10^{-6}$, and median $\sim 10^{-10}$.
    }
    \label{fig:state_prep}
\end{figure}

The stabilization timescale $\tau_s$ under small perturbations around the target scales like $\tau_s \sim \left|\Delta \gamma \cos(\theta) + 2 (\gamma_+ +\gamma_-) \sin^2(\theta) \cos^2(\phi)\right|^{-1}$.
Thus, stabilization is generally more effective for larger damping gaps $\Delta \gamma$.
This means that, to accelerate convergence in the upper hemisphere, one must increase the pumping rate $\gamma_+$.
For the lower hemisphere, one should set $\gamma_+ = 0$, thereby decoupling the energy reservoir from the system.
This is illustrated in Fig.~\ref{fig:state_prep}.
We prepare a state in the upper hemisphere with Bloch angles $(\theta,\, \phi) = (\pi/4,\, \pi/4)$, for an initial condition equal to $\ket{1}$.
By considering a non-negligible gap $\Delta \gamma$, we achieve a relatively fast convergence and stabilization around the target of an ensemble of quantum trajectories.

\subsubsection{Example 2: entanglement steering of a two-qubit system}
\label{sec:entanglement_stabilization}

Consider the parametrized target state $Q_\alpha = \ket{q_\alpha} \bra{q_\alpha}$, where $\ket{q_\alpha}$ is defined in Schmidt form as $\ket{q_\alpha} = \sqrt{\alpha} \ket{00} + \sqrt{1-\alpha}\ket{11}$ for $\alpha \in [0, \, 1]$.
This state interpolates between product states at $\alpha = 0$ and $\alpha = 1$, and the Bell state $\ket{\Phi^+} = (\ket{00} + \ket{11})/\sqrt{2}$ at $\alpha = 1/2$, which is maximally entangled according to, e.g., the von Neumann entropy measure~\cite{nielsen2010quantum}.

We define two measurement channels through the jump operators $C_1 = X_1 (1 - Z_1Z_2)/2$ and $C_2 = Z_1 (1 - X_1 X_2)/2$, where $X_i$ and $Z_i$ are the Pauli $x$ and $z$ operators, respectively, acting on qubit $i$.
A purely dissipative Lindblad dynamics driven by $C_1$ and $C_2$ has the Bell state $\ket{\Phi^+}$ as its unique steady state.
When all measurement outcomes are discarded and no feedback is applied, the evolution governed by the Lindblad equation converges to $\ket{\Phi^+}$.
Ref.~\cite{barreiroOpenSystemQuantumSimulator2011} exploits this mechanism to achieve autonomous dissipative quantum state preparation, where a two-qubit system is cooled into one of the four Bell states.
The joint effect of the operators $C_1$ and $C_2$ on the populations of the Bell states is as follows.
The four Bell states, $\ket{\Phi^\pm} = (\ket{00} \pm \ket{11})/\sqrt{2}$ and $\ket{\Psi^\pm} = (\ket{01} \pm \ket{10})/\sqrt{2}$, are simultaneous eigenstates of $Z_1 Z_2$ and $X_1X_2$, with eigenvalues $\pm 1$.
Consequently, the jump operators $C_1$ and $C_2$ collectively pump population from the three Bell states orthogonal to $\ket{\Phi^+}$ into $\ket{\Phi^+}$, while $\ket{\Phi^+}$ remains a dark state.
More specifically, they map the $-1$ eigenspaces of $Z_1 Z_2$ and $X_1 X_2$ into the corresponding $+1$ eigenspaces.
For instance, $C_1$ maps $\ket{\Psi^-}$ to $-\ket{\Phi^-}$, and $C_2$ maps $\ket{\Phi^-}$ to $\ket{\Phi^+}$.

By continuously monitoring the system and incorporating measurement-based feedback into its dynamics, we can alter the prior behavior and steer the fixed point of the evolution toward a different entangled state $Q_\alpha \neq \ket{\Phi^+}\bra{\Phi^+}$.
We do this by implementing the state-agnostic feedback protocol outlined in Eq.~\cref{eq:sme_state_agnostic_fb} for the two dissipative channels $C_1$ and $C_2$.
The feedback dynamics is
\bal{\label{eq:entanglement_fb}
d\rho_t =&
        \frac{\gamma_1}{4} [[Z_1 Z_2, \rho_t], \rho_t] dt  \nonumber\\
        &+\frac{\gamma_2}{4} [[X_1 X_2, \rho_t - Q_\alpha], \rho_t] dt \nonumber\\
        &+ \frac{1}{2}[[X_1, \rho_t - Q_\alpha], \rho_t]dy_t^{(1)} \nonumber\\
        &+ \frac{1}{2}[[Z_1, \rho_t - Q_\alpha], \rho_t]dy_t^{(2)}\,.
}
The measurement outcomes are $dy_t^{(1)} = \gamma_1 \braket{X_1}_t dt + dW_t^{(1)}$ and $dy_t^{(2)} = \gamma_2 \braket{Z_1}_t dt + dW_t^{(2)}$, with $\big(dW_t^{(1)}\big)^2 = \gamma_1 dt$ and $\big(dW_t^{(2)}\big)^2 = \gamma_2 dt$, where $\gamma_1$ and $\gamma_2$ are the damping rates for $C_1$ and $C_2$, respectively.
Equation~\cref{eq:entanglement_fb} is obtained from~\cref{eq:sme_state_agnostic_fb} by taking $A_1 = \frac{X_1}{2}$ and $B_1 = \frac{i}{2}Y_1 Z_2$, which gives $[A_1, B_1] = -\frac{1}{2}Z_1 Z_2$, and $A_2 = \frac{Z_1}{2}$ and $B_2 = -\frac{i}{2}Y_1 X_2$, which gives $[A_2, B_2] = -\frac{1}{2}X_1 X_2$.
We also use the fact that $A_1^2=A_2^2 \propto I$ to eliminate the corresponding drift terms, and that $Q_\alpha$ commutes with $Z_1 Z_2$.

When $\alpha=1/2$, $Q_\alpha = \ket{\Phi^+}\bra{\Phi^+}$ commutes with $X_1 X_2$, so the drift term in~\cref{eq:entanglement_fb} becomes a pure double-bracket term driving the system towards the $+1$ sectors of the eigenspaces of $Z_1 Z_2$ and $X_1 X_2$.
In other words, we recover state stabilization for the Bell state $\ket{\phi^+}$, albeit with active feedback.

When $\alpha \neq 1/2$, the stable fixed-point of the dynamics is tilted toward $Q_\alpha$.
We establish the stability of $Q_\alpha$ by applying the Lyapunov stability analysis described in App.~\ref{app:stability_analysis}.
The target state $Q_\alpha$ with $\alpha = 1/2$ is stable provided that
\bal{\label{eq:entanglement_stability}
-\gamma_1 (1 - \braket{Z_1Z_2}_\perp) -\gamma_2 (1 - \braket{X_1 X_2}_\perp) < 0\,,
}
where $\braket{\cdot}_\perp$ denotes the expectation value with respect to a normalized state orthogonal to $\ket{q_\alpha}$.
Eq.~\cref{eq:entanglement_stability} follows using the definitions of $C_1$ and $C_2$ in the general stability condition~\cref{eq:app_lyapunov_general} given in App.~\ref{app:stability_analysis}.
Since relation~\cref{eq:entanglement_stability} holds true for  any orthogonal state to $\ket{q_\alpha}$, the $Q_\alpha$ is a stable fixed-point of the dynamics.
When $\lambda_1 = \lambda_2$, a similar argument holds true for arbitrary $\alpha \in [0, 1]$, which we show in App.~\ref{app:stab_entanglement}.

\begin{figure}
    \centering
\includegraphics[width=0.9\columnwidth]{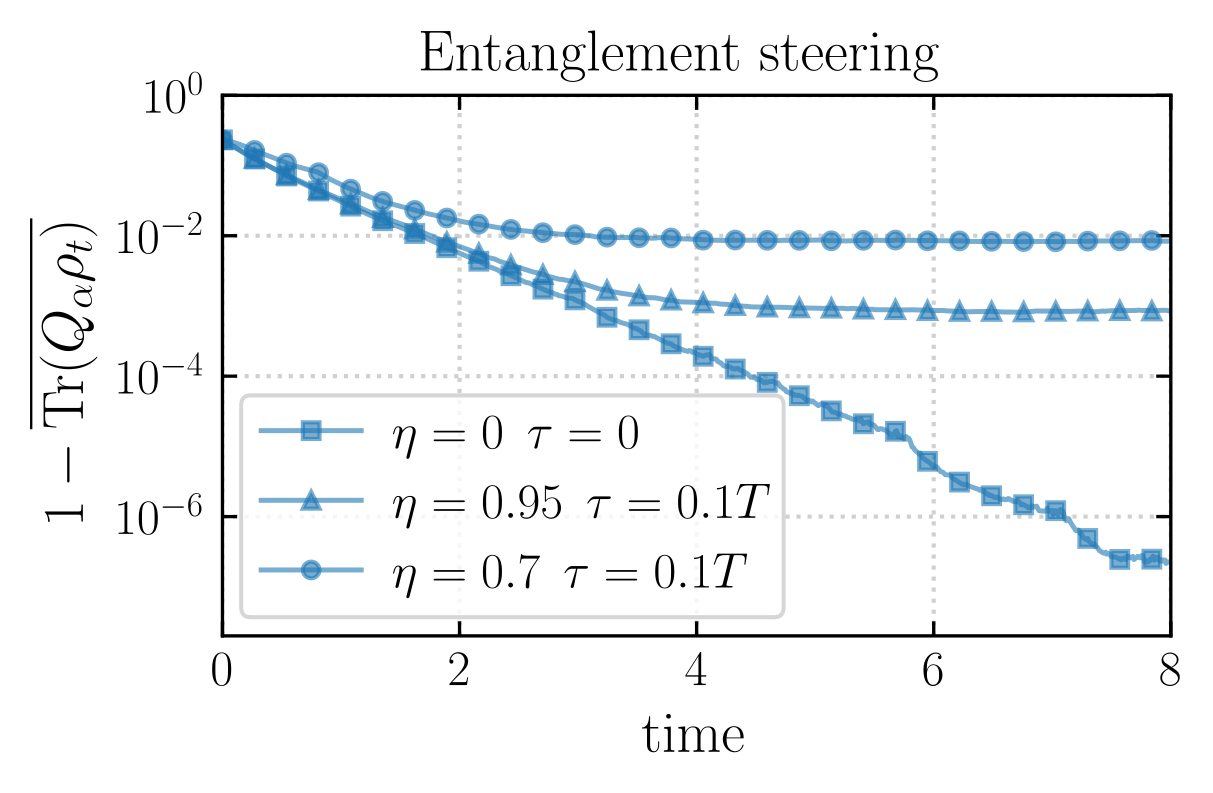}
    \caption{
    Entanglement steering and stabilization of a two-qubit system with dynamics given by~\cref{eq:entanglement_fb}.
    We compute the average infidelity $1 - \overline{\tr{Q_\alpha\rho_t}}$ over time for a single random perturbation around the target $Q_{\alpha}$ with $\alpha = 0.75$.
    The parameters are: final time $T=8$ with $800$ time steps, damping rates $\gamma_1 = \gamma_2 = 3$, and $10000$ trajectories per average infidelity curve.
    In the ideal case (squares) with $\eta=1$ and $\tau=0$ the average infidelity decays exponentially fast, reaching values $\sim 10^{-7}$.
    In the non-ideal cases, the exponential decay slows down and plateaus at higher final values.
    For mild imperfect measurements with $\eta=0.95$ and delay $\tau = 0.1 T$ (triangles), the final average infidelity is $\sim 10^{-3}$.
    For a lower measurement efficiency $\eta=0.7$ and delay $\tau = 0.1T$ (circles), a regime that is experimentally relevant in superconducting-qubit platforms~\cite{eddins2019highefficiency, lecocq2021efficient}, the final average infidelity is $\sim 10^{-2}$.
    We validated these results across multiple random perturbations of the target state, obtaining similar convergence behavior.
    }  
\label{fig:entanglement}
\end{figure}

In practice, non-idealities such as imperfect measurements and feedback delay are expected to partially degrade the performance of the stabilization protocol relative to the ideal case.
Imperfect measurements are characterized by an efficiency parameter $0 \le \eta \le 1$, which represents the fraction of measurement outcomes that are detected and incorporated into the feedback loop, whereas feedback delay refers to applying at time $t$ a feedback Hamiltonian constructed from records acquired at time $t-\tau$.
In Fig.~\ref{fig:entanglement}, we contrast the ideal and non-ideal cases for the target state $\alpha = 0.75$.
For a random perturbation around the target $Q_\alpha$, we plot the average infidelity $1- \overline{\tr{Q \rho_t}}$ over time.
In the ideal case, with $\eta= 1$ and $\tau=0$, the average infidelity decays exponentially in time, as expected from the double-bracket structure of the dynamics.
In the non-ideal cases, a similar exponential decay is observed at short times, followed by saturation at a higher infidelities than in the ideal scenario.

%================================================================================

\vspace{7pt}
\subsection{Gradient flow interpretation of feedback dynamics}\label{sec:GradientFlowFeedback}

Next, we interpret general state-agnostic feedback protocols within the context of gradient flows.
A general feedback of this kind can be implemented by adding a Hamiltonian generator $dH_t^\fb$ to the dynamics~\cref{eq:sme_strato}, so that the latter is transformed into
\bal{\label{eq:fb_gen}
d\rho_t &=  -i\cC_\rho (dH_t^\sb) + \cC_\rho^2 (dH_t^\db) -i \cC_\rho(dH_t^\fb)\,.
}
Using the periodic property $\cC_\rho^3 = \cC_\rho$, we can write the feedback equation in DB form
\bal{\label{eq:db_fb}
d\rho_t &=  \cC_\rho^2 \lpar 
-i\cC_\rho (dH_t^\sb + dH_t^\fb) + dH_t^\db
\rpar\,.
}
As a direct application of Prop.~\ref{prop:stoch_flow} to Eq.~\cref{eq:db_fb} gives the following.
\begin{coro}
\label{coro:fb}
The feedback dynamics~\cref{eq:fb_gen} is equivalent to a gradient flow $d\rho_t = \nabla F_t(\rho_t)dt$ with potential
\bal{
F_t(\sigma) = \tr{\lpar -i\cC_\rho (\dot H_t^\sb + \dot H_t^\fb) + \dot H_t^\db \rpar \sigma}\,.
}
\end{coro}

Enforcing fixed-point dynamics would demand the design of a feedback process such that $\nabla F_t(\rho_t = \text{fixed point}) = 0$.
This is indeed the case for the state-agnostic protocol treated in the previous section, where the target is an arbitrary state $Q=\ket{q}\!\bra{q}$.
In that case, the feedback process is given by $dH_t^\fb = -dH_t^\sb -i\cC_Q(dH_t^\db)$, where $\cC_Q(\bullet) = [\bullet,\, Q]$, and we write the dynamics as
\baln{
d\rho_t = \cC_\rho \cC_{\rho - Q}(dH_t^\db)\,.
}
When $\rho = Q$, $d\rho_t \equiv 0$, as expected.
The corresponding potential is $F_t(\sigma) = \tr{\cC_\rho \cC_{\rho - Q}(\dot H_t^\db)\sigma}$, with flow equation
\baln{
d\rho_t = -\gamma \nabla F^{(1)}(\rho_t) dt + \nabla F^{(2)}(\rho_t) dy_t\,,
}
where
\baln{
F_t^{(1)}(\sigma) \coloneqq \tr{\cC_\rho \cC_{\rho - Q}(A^2 + [A,\, B]/2)\sigma}
}
and
\bal{\label{eq:F2}
F_t^{(2)}(\sigma) \coloneqq \tr{\cC_\rho \cC_{\rho - Q}(A)\sigma}\,.
}

In the previous section, we showed that a Lyapunov stability analysis determines whether the solution $\rho = Q$ is a stable fixed point and provides sufficient conditions under which this is achieved.
With Corollary~\ref{coro:fb}, we can re-interpret the stability of $Q$ as that of a stationary solution of a certain loss landscape.
For this, we compute the change in the potential increment
\bal{\label{eq:fb_potential_increment}
F_t(\sigma) dt = \tr{\cC_\rho \cC_{\rho - Q}(dH_t^\text{DB})\sigma}
}
under a small dynamical update $\rho_t \goto \rho_t + d \rho_t$ in the direction of the flow.
The result is
\baln{
(F_t dt) (\rho_t + d\rho_t) = \gamma \big\| \nabla F^{(2)}_t(\rho_t) \big\|_\cO^2 dt + \bigo{dt^2}\,.
}
This shows that the potential increment $F_t dt$ is non-decreasing when the state is updated according to the flow.
Note that, in particular, for $\rho = Q$, the gradient vanishes $\nabla F^{(2)}_t(Q) = 0$, which implies that the potential has reached a stationary solution.
Whether the solution is stable or not depends on a second order analysis of $F_t(\sigma) dt$ around $\sigma = Q$.
An approach is the following.

Note that, in general, the increment~\cref{eq:fb_potential_increment} quantifies the change in the potential when we move along the orbit of $\rho_t$.
Since the increment is non-negative, if $Q$ is a stable fixed-point, we must have
\bal{\label{eq:cond_increment}
F_t(\rho_t + \tau \delta \rho)dt  = F_t(Q)dt > 0\,,
}
for any non-zero difference $Q - \rho_t = \tau \delta \rho$, with $\tau > 0$ sufficiently small and $\delta\rho$ being any fixed traceless operator such that $\rho_t$ remains in the orbit of $Q$.
Because
\baln{
F_t(Q) dt = \tau \underbrace{
\tr{Q \cC_Q \cC_{\delta \rho}(dH_t^\db)}
}_{= 0} + \tau^2 \tr{Q \cC_{\delta \rho}^2(dH_t^\db)}\,,
}
condition~\cref{eq:cond_increment} becomes $\tr{Q \cC_{\delta \rho}^2(dH_t^\db)} > 0$.
We can satisfy this in expectation if
\baln{
\expec{\tr{Q \cC_{\delta \rho}^2(d H_t^\db)}} > 0\,,
}
where the average $\expec{\cdot}$ is over all noise realizations.
After replacing $dH_t^\db$ by its definition, the condition above becomes
\bal{\label{eq:local_extrema}
\tr{Q \cC_{\delta \rho}^2 \lkey A^2 + [A, B]/2 - 2\tr{QA} A \rkey} < 0
}
for all traceless $\delta\rho$ such that $\rho$ is in the orbit of $Q$.

Since the diffusion potential $F^{(2)}_t(\sigma)$ in~\cref{eq:F2} tends to vanish as the state approaches the target $Q$, the variance in the trajectories becomes increasingly suppressed, increasing the chances that individual trajectories are trapped in the surroundings of $Q$.
Condition~\cref{eq:local_extrema} is equivalent to the stability condition that one derives in the context of Lyapunov stability (see App.~\ref{app:stability_analysis}).

The addition of the state-agnostic feedback with target $Q$ introduces an additional fixed-point solution to the landscape of the potential, but at the same time it modifies it.
For instance, when $B=0$ and in the absence of feedback ($Q \equiv 0$), the stable fixed points of the dynamics correspond to the eigenspaces of $A$.
If we switch on the feedback ($Q \neq 0$), we introduce a new fixed point equal to $Q$, but we also remove the eigenspaces of $A$ from the set of fixed points unless $[Q, A]= 0$.
This indicates that the effect of the feedback process in the potential landscape is to tilt the fixed points in favor of the target $Q$.

%================================================================================

\vspace{7pt}
\section{Discussion and Outlook}
\label{sec:discussion}

There tends to be a tension between the unitary evolution of isolated quantum systems and measurement processes. The former describes smooth, differentiable dynamics, with a Hamiltonian well-determined by fields and interactions. 
In contrast, quantum measurements are typically presented as discontinuous jumps that break from Hamiltonian dynamics. 
Our results provide an alternative to the latter picture, one in which measurements are also described by stochastic Hamiltonians that, in turn, lead to double-bracket gradient flows. 
In this way, we unify double-bracket gradient flows, measurement processes, and Hamiltonians.

One could thus conceive of a picture where the collapse postulate is replaced by a `measurement Hamiltonian postulate', in which measurements are described by $\Hsb$ and $\Hdb$ in Eqs.~\eqref{eq:Hsb} and~\eqref{eq:Hdb}, with a stochastic measurement output as in Eq.~\eqref{eq:dyt}. Alternatively, one could assume a gradient flow with the stochastic potentials defined in Eqs.~\eqref{eq:potentialSB} and~\eqref{eq:potentialDB}.

Single- and double-bracket gradient flows describe the ensuing dynamics. 
In the pure-measurement case ($B = 0$ and $\|H_0\| \ll \{\|H^\sb\|,\| H^\db \|\}$), the coarse-grained dynamics and statistics of measurement outcomes match the Born rule. 
Naturally, such approaches would still entail an uncomfortable pitfall shared with collapse-based interpretations: nothing in the theory says when a measurement occurs~\cite{bell1990against} (i.e., when dynamics is dictated by $H^\sb + H^\db$ rather than $H_0$). 
In this respect, interpretations that leverage environmental decoherence to explain measurement occurrences benefit from criteria for defining when a measurement occurs~\cite{Schlosshauer}. 
However, interpretations based on environmental decoherence often suffer from the ``\emph{and/or problem}'', as no definite outcome is selected~\cite{bell1990against, Schlosshauer}. In contrast, the stochastic measurement Hamiltonians introduced in this paper produce a single (random) outcome, the minimum of a gradient flow.
 
We have leveraged the generalized framework to describe measurement processes that inform novel quantum feedback protocols. In particular, we provided a feedback mechanism that results in deterministic dynamics for a double-bracket flow form. In such a process, the system ends up in the ground state of a Hamiltonian. We also showed a state-agnostic feedback process that can prepare arbitrary states of a qubit.
More generally, the framework allows interpreting feedback processes for state preparation as  gradient flows with a feedback-modified stable fixed point. 
Future studies examining the interplay between Lyapunov stability, non-equilibrium physics, and feedback control may offer deeper insight into the mechanisms that drive collapse phenomena.

We presented an example of entanglement stabilization exhibiting exponential convergence in the ideal case.
We further showed, through this example, that the introduction of non-idealities such as measurement inefficiency and feedback delay does not dramatically degrade performance relative to the ideal case within the range of current experimental capabilities, while still enabling the stabilization of entangled states with high precision.
Experimental realizations of feedback protocols that explicitly account for such non-idealities constitute an important direction for future work.

It is natural to draw conceptual parallels between imaginary-time evolution constructions and the monitored double-bracket dynamics studied here.
An important avenue for future research is the exploration of measurement-based variants of double-bracket quantum algorithms~\cite{xiaoyueStrategiesOptimizingDoublebracket2024,gluzaDoublebracketQuantumAlgorithms2024, suzuki2025double, gluzaDoublebracketQuantumAlgorithms2025}, grounded in the monitored double-bracket dynamics studied in this work.

%================================================================================

\vspace{7pt}
\section*{Acknowledgments}
A.V. acknowledges support from the U.S. Department of Energy (DOE) through a quantum computing program sponsored by the Los Alamos National Laboratory (LANL) Information Science $\&$ Technology Institute.
This material is based upon work supported by the U.S. DOE, Office of Science, National Quantum Information Science Research Centers, Quantum Science Center, 
and the Beyond Moore’s Law project of the Advanced Simulation and Computing Program at LANL.

%================================================================================

\vspace{7pt}
\section*{Data availability}

The data that support the findings of this study are available from the corresponding author upon reasonable request.

\bibliography{references}

%================================================================================
\onecolumngrid

\appendix

\section{Monitored dynamics in the Itô picture}\label{app:ito}

We provide a self-contained derivation of the SME~\cref{eq:sme_ito} for pure states, i.e., when the density operator satisfies $\rho^2=\rho$.
The standard non-linear stochastic Schrödinger equation (SSE) for a single noise channel with jump operator $C$ is given by
\bal{\label{eq:dpsi_cond_uncontrolled_ito}
    d\psi_t = -i H_e(t) \psi_t dt + (C - m_t) \psi_t dW_t\,,
}
where $dW_t$ is a zero-mean Wiener increment such that $dW_t^2 = \gamma  dt$ with $\gamma > 0$~\cite{barchielliQuantumTrajectoriesMeasurements2009}.
We define the parameter $m_t \coloneqq \psi^\dagger_t C^\H \psi_t = \braket{C^\H}_t$, with $C^\H = (C+ C^\dagger)/2$ the Hermitian part of $C$, and the time-dependent operator
\baln{
    H_e(t) \coloneqq H_0(t) -\frac{i}{2} \gamma(C^\dagger C - 2 m_t C + m_t^2)\,.
}
Note that $H_e$ is not Hermitian.
Equation~\cref{eq:dpsi_cond_uncontrolled_ito} satisfies $d\|\psi_t\|^2 = 0$, i.e., the norm of $\psi_t$ is conserved deterministically.
Equation~\cref{eq:dpsi_cond_uncontrolled_ito} represents the dynamics of the state $\psi_t$ conditioned on the measurement process $y_t$~\cite{barchielliQuantumTrajectoriesMeasurements2009}, which is characterized by the measurement output
\baln{
dy_t = 2 \gamma m_t dt + dW_t\,.
}

We compute the SME for the density operator $\rho_t = \psi_t \psi_t^\dagger$ by applying the Itô rule~\cite{oksendal2003stochastic}:
\bal{\label{eq:dpsidpsi}
d\rho_t = d(\psi_t \psi_t^\dagger) = (d\psi_t) \psi_t^\dagger + \psi_t (d\psi_t^\dagger) + (d\psi_t) (d\psi_t^\dagger)\,.
}
From the SSE~\cref{eq:dpsi_cond_uncontrolled_ito}, we compute the first two terms in~\cref{eq:dpsidpsi} as
\baln{
    (d\psi_t)\psi_t^\dagger &= -iH_e(t)\psi_t\psi_t^\dagger\,dt + (C - m_t)\psi_t\psi_t^\dagger\,dW_t \notag \\
    &= -iH_e(t)\rho_t\,dt + (C - m_t)\rho_t\,dW_t, \\[6pt]
    \psi_t(d\psi_t^\dagger) &= i\psi_t\psi_t^\dagger H_e^\dagger(t)\,dt + \psi_t\psi_t^\dagger(C^\dagger - m_t)\,dW_t \notag \\
    &= i\rho_t H_e^\dagger(t)\,dt + \rho_t(C^\dagger - m_t)\,dW_t\,,
}
where we used that $m_t = \psi_t^\dagger C^{\mathrm{H}} \psi_t \in \mathbb{R}$ is real, so the complex conjugate $m_t^* = m_t$, and the definition of $\rho_t = \psi_t \psi_t^\dagger$.

For the third term in~\cref{eq:dpsidpsi}, only the stochastic terms survive (since $dt dW_t = 0$ and $(dt)^2 = 0$), and using $dW_t^2 = \gamma\,dt$ we obtain:
\bal{
    (d\psi_t)(d\psi_t^\dagger)
    &= \bigl((C - m_t)\psi_t\,dW_t\bigr)\bigl(\psi_t^\dagger(C^\dagger - m_t)\,dW_t\bigr) \notag \\
    &= (C - m_t)\psi_t\psi_t^\dagger(C^\dagger - m_t)\,dW_t^2 \notag \\
    &= \gamma(C - m_t)\rho_t(C^\dagger - m_t)\,dt\,.
}

Combining all terms, we obtain
\bal{\label{eq:drho_1}
    d\rho_t = -i (H_e(t) \rho_t - \rho_t H_e^\dagger(t)) dt + \gamma(C - m_t) \rho_t (C^\dagger - m_t) dt + \lpar (C-m_t)\rho_t + \rho_t(C^\dagger - m_t) \rpar dW_t\,.
}

We expand the drift term in~\cref{eq:drho_1} as
\baln{
-i (H_e(t) \rho_t - \rho_t H_e^\dagger(t)) dt + \gamma(C - m_t) \rho_t (C^\dagger - m_t) dt =& -i[H_0,\rho_t]dt 
+\gamma\Big\{
        (C-m_t)\rho_t(C^\dagger-m_t) \nonumber \\
        =& -\tfrac{1}{2}(C^\dagger C - 2m_t C + m_t^2)\rho_t
        -\tfrac{1}{2}\rho_t(C^\dagger C - 2m_t C^\dagger + m_t^2) \Big\} dt \nonumber\\
        =& -i[H_0,\rho_t]dt + \gamma \lpar C\rho_tC^\dagger - \frac{1}{2} \lkey C^\dagger C, \rho_t \rkey \rpar\,.
}
In the last equality, we recover the Lindbladian $\cL(\rho_t) = -i[H_0, \rho_t] + \cD[C](\rho_t)$, with dissipation part $\cD[C](\rho_t) = \gamma \lpar C\rho_tC^\dagger - \frac{1}{2} \lkey C^\dagger C, \rho_t \rkey \rpar$.
By considering the conjugate anticommutator between operators $X, Y$ given by $\{X, Y\}_c \coloneqq XY + Y^\dagger X^\dagger$ (which is just twice the Hermitian part of $XY$), we arrive at the desired form of~\cref{eq:drho_1}:
\bal{\label{eq:dP_unraveling}
    d\rho_t = \cL(\rho_t) dt + \{C-m_t, \rho_t\}_c dW_t\,,
}
which corresponds to~\cref{eq:sme_general_ito} in the main text by identifying $m_t = \braket{A}_t$ where $A = C^\H$.

To derive~\cref{eq:sme_ito} in the main text, we recast the diffusion coefficient in~\cref{eq:dP_unraveling} as a commutator by splitting the jump operator $C$ in Hermitian $A = C^\H$ and anti-Hermitian $B = C^\AH$ parts:
\bal{\label{eq:conj_anticomm}
    \{C - m_t, \rho_t\}_c =& C\rho_t + \rho_t C^\dagger - 2m_t\rho_t \nonumber\\
    =& (C^\H+C^\AH)\rho_t + \rho_t(C^\H-C^\AH) - 2m_t\rho_t \nonumber \\
    =&C^\H\rho_t + \rho_t C^\H + C^\AH\rho_t - \rho_t C^\AH - 2\rho_t C^\H\rho_t \nonumber \\
    =& [C^\H,\rho_t]\rho_t - \rho_t[C^\H,\rho_t] + [C^\AH,\rho_t] \nonumber \\
    =&[[A, \rho_t] + B, \rho_t]\,,
}
where in the third line we used that $\rho_t C^\H \rho_t = m_t \rho_t$, in the fourth line we used $\rho_t^2 = \rho_t$,
and in the last line we used the definitions of $A$ and $B$.
Defining the stochastic Hamiltonian $H_t \coloneqq i([A, \rho_t] + B)$ we use the equivalence~\cref{eq:conj_anticomm} to finally write~\cref{eq:dP_unraveling} as
\bal{\label{eq:app_sme_single}
    d\rho_t = \cL(\rho_t) dt -i [H_t, \rho_t] dW_t\,,
}
which is~\cref{eq:sme_ito} in the main text.

The generalization of~\cref{eq:app_sme_single} to $n_c$ multiple noise channels is straightforward.
Consider $n_c$ Wiener processes $W_a(t)\,(a=1,\,\ldots,\, n_c)$ such that $dW_a(t) dW_b(t) = \delta_{ab} \gamma_a dt$ with $\gamma_a > 0$, and define $n_c$ arbitrary jump operators $C_a$.
The SSE is
\baln{
    d\psi_t = -i H_e(t) \psi_t dt + \sum_a (C_a - m_a(t)) \psi_t dW_a(t)\,,
}
where $H_e(t) = H_0(t) -\frac{i}{2} \sum_a \gamma_a(C_a^\dagger C_a - 2 m_a(t) C_a + m_a^2(t))$.
Then, the unraveling for the state $\rho_t$ is given by
\baln{
    d\rho_t &= \cL(\rho_t) dt + \sum_a \{C_a-m_a(t), \rho_t\}_c dW_a(t)
}
where $\cL(\rho_t) = -i[H_0, \rho_t] +  \sum_a \cD_a(\rho_t)$ with $\cD_a(\rho_t) = \gamma_a \lpar C_a \rho_t C_a^\dagger - \frac{1}{2} \lkey C_a^\dagger C_a, \rho_t \rkey \rpar$, and $m_a(t) = \psi^\dagger_t C^\H_a \psi_t$.

%==============================================================================================

\section{Itô-Stratonovich conversion formula}\label{app:ito-strato}

Consider the following complex-valued SDE written in Stratonovich form:
\bal{\label{eq:strato_prod}
    dY_a(t) = f(t, X_t) \circ dW_a(t)\,,
}
where $dW_a(t)\,(a=1,\ldots, n_c)$ are real Wiener increments satisfying $dW_a(t) dW_b(t) = \Gamma_{ab} dt$ with $\Gamma > 0$ a real symmetric matrix.
The process $X_t \in \complex^N$ satisfies $dX_t = \mu(t, X_t) dt + \sigma(t, X_t) dW_t$ with $\mu: [0, T]\times \complex^N \goto \complex^N$ and $\sigma: [0, T]\times \complex^N \goto \complex^{N \times n_c}$, and $dW_t$ is the vector of Wiener increments.
We assume $\sigma$ is smooth, so it admits a Taylor expansion.
The symbol $\circ$ in~\cref{eq:strato_prod} denotes mid-point evaluation of $f(t, X_t)$ and thus specifies the Stratonovich interpretation of the stochastic integral~\cite{oksendal2003stochastic,kloeden2012numerical}, whereby
\bal{\label{eq:strato_def}
f(t, X_t) \circ dW_a(t) \equiv f(t + dt/2, X_t + dX_t/2) dW_a(t)\,,
}
while the second term in~\cref{eq:strato_def} is written in the It\^o convention.
We can derive the Itô form of~\cref{eq:strato_prod} by applying the definition of mid-point evaluation followed by a stochastic Taylor expansion of $f$~\cite{kloeden2012numerical}:
\baln{
    dY_a(t) &= f(t, X_t) \circ dW_a(t) \nonumber\\
    &= f(t + dt/2, X_t + dX_t/2) dW_a(t) \nonumber\\
    &= \lpar f(t, X_t) + \frac{1}{2}\partial_t f (t, X_t) dt + \frac{1}{2}\partial_b f(t, X_t) dX_t^b + \ldots \rpar dW_a(t) \nonumber\\
    &= f(t, X_t) dW_a +\frac{1}{2} \partial_b f(t, X_t) dX_t^b dW_a \nonumber\\
    &= f(t, X_t) dW_a(t) + \frac{1}{2}\partial_b f(t, X_t) \sigma^{b c}(t, X_t) dW_c(t) dW_a(t)
}
Using $dW_c(t) dW_a(t) = \Gamma_{ca}dt$, we obtain the desired Itô-Stratonovich transformation in vector-matrix form
\bal{\label{eq:ito-Strato}
    f(t, X_t) \circ dW(t) = f(t, X_t) dW(t) + \frac{1}{2} \nabla f(t, X_t) \sigma_t(t, X_t) \Gamma dt\,,
}
where $\nabla f$ is the gradient of $f$.

%================================================================================
\section{Proof of the periodic property of $\cC_\rho$}
\label{app:periodic_property}

For a pure state $\rho$ with $\rho^2 = \rho$, the super-operator $\cC_\rho(\bullet) = [\bullet, \rho]$ satisfies:
\baln{
\cC_\rho^2(X) 
  &= [[X, \rho], \rho] 
   = [X\rho - \rho X,\, \rho] \notag \\
  &= X\rho^2 - \rho X\rho - \rho X\rho + \rho^2 X 
   = X\rho - 2\rho X\rho + \rho X\,,
}
where $X$ is an arbitrary operator.
Applying $\cC_\rho$ once more and expanding using $\rho^2 = \rho$:
\baln{
\cC_\rho^3(X) 
  &= [X\rho - 2\rho X\rho + \rho X,\, \rho] \notag \\
  &= X\rho^2 - \rho X\rho - 2\rho X\rho^2 + 2\rho^2 X\rho 
     + \rho X \rho - \rho^2 X \notag \\
  &= X\rho - \rho X = \cC_\rho(X).
}
Hence $\cC_\rho^3 = \cC_\rho$, and by induction 
$\cC_\rho^{2k+1} = \cC_\rho$ for all $k \geq 0$.
This further implies that $\cC_\rho^4 = \cC_\rho^3 \cC_\rho = \cC_\rho \cC_\rho = \cC_\rho^2$, and by induction $\cC_\rho^{2k} = \cC_\rho^2$ for all $k \ge 1$.

%================================================================================

\section{Proof of Theorem~\ref{th:sme_strato}}
\label{app:theorem_1}
We write the proof of Theorem~\ref{th:sme_strato} in two steps.
In the first step, we transform the SME~\cref{eq:dP_unraveling} from It\^o to the Stratonovich picture.
In the second step, we derive the Hamiltonian form of the resulting dynamics.

\subsection{Monitored dynamics in the Stratonovich picture}\label{app:strato}

To transform the SME~\cref{eq:dP_unraveling} into the Stratonovich picture, we first rewrite~\cref{eq:dP_unraveling} as 
\bal{\label{eq:sme_ito_sigma}
d\rho_t = \cL(\rho_t) dt + \sigma_t dW_t
}
where, for notational convenience, we define $\sigma_t \coloneqq \{C-m_t, \rho_t\}_c$.
Note that $\sigma_t$ depends on $\psi_t$ only through $\rho_t$, since $m_t = \braket{C^\H}_t \equiv \tr{C^\H \rho_t}$.

We use the Itô-Stratonovich transformation from App.~\ref{app:ito-strato}, instantiated for the present setup with $f = \sigma_t$ and $\Gamma$ one-dimensional with $\Gamma = \gamma$, to write the diffusion term in~\cref{eq:sme_ito_sigma} component-wise as 
\bal{\label{eq:ito-Strato-particular}
    \sigma_t^{ab} dW_t = \sigma_t^{ab} \circ dW_t - \frac{\gamma}{2} \sigma_t^{cd} \partial_{cd} \sigma_t^{ab} dt\,,
}
where the derivatives $\partial_{cd}$ are with respect to the coordinates $(\rho_t)_{cd}$, and $\circ$ denotes the mid-point evaluation rule in the Stratonovich picture.
Repeated indices denote summation.
Note that the contraction $\sigma_t^{cd}\partial_{cd}\sigma_t$ can be written as the application of the linear operator $\partial_\rho \sigma_t$ on $\sigma_t$, where the partial derivative is taken component-wise.
Then,
\bal{\label{eq:partial_sigma}
    \sigma_t^{cd}\partial_{cd}\sigma_t = (\partial_\rho \sigma_t) (\sigma_t) =& (C - m_t) \sigma_t + \sigma_t (C^\dagger - m_t) - 2 \tr{C^\H \sigma_t} \rho_t\,.
}
We expand the first two terms as
\baln{
(C-m_t)\sigma_t + \sigma_t(C^\dagger-m_t)
&= C\sigma_t + \sigma_t C^\dagger - 2m_t\sigma_t \\
&= C(\{C,\rho_t\}_c - 2m_t\rho_t)
   + (\{C,\rho_t\}_c - 2m_t\rho_t)C^\dagger
   - 2m_t(\{C,\rho_t\}_c - 2m_t\rho_t) \\
&= C(C\rho_t + \rho_t C^\dagger - 2m_t\rho_t)
 + (C\rho_t + \rho_t C^\dagger - 2m_t\rho_t)C^\dagger
 - 2m_t(C\rho_t + \rho_t C^\dagger - 2m_t\rho_t) \\
&= C^2\rho_t + \rho_t(C^\dagger)^2 + 2C\rho_t C^\dagger
 - 4m_t(C\rho_t + \rho_t C^\dagger) + 4m_t^2\rho_t \\
&= \{C^2,\rho_t\}_c - 4m_t\{C,\rho_t\}_c
 + 2C\rho_t C^\dagger + 4m_t^2\rho_t\,,
}
where, in the second line, we used that $\sigma_t = \{ C, \rho_t\}_c - 2m_t \rho_t$.

For the trace term,
\baln{
\operatorname{Tr}(C^H\sigma_t)
&= \operatorname{Tr}\!\left[
C^H\bigl((C-m_t)\rho_t + \rho_t(C^\dagger-m_t)\bigr)
\right]
\\
&= \operatorname{Tr}(C^HC\,\rho_t)
 + \operatorname{Tr}(C^H\rho_t C^\dagger)
 - 2m_t\operatorname{Tr}(C^H\rho_t)
\\
&= \langle C^HC\rangle_t + \langle C^\dagger C^H\rangle_t - 2m_t^2
\\
&= \bigl\langle \{C^H,C\}_c \bigr\rangle_t - 2m_t^2\,,
}
where, in the third line, we used the cyclic property of the trace and the definition of $m_t = \tr{C^\H \rho_t}$.
Replacing back into~\cref{eq:partial_sigma} we obtain
\bal{\label{eq:partial_sigma_2}
(\partial_\rho \sigma_t) (\sigma_t) =& \{C^2, \rho_t \}_c - 4 m_t \{C, \rho_t\}_c + 2 C\rho_tC^\dagger + 8m_t^2 \rho_t - 2 \braket{\{ C^\H, C \}_c}_t \rho_t
}

Inserting~\cref{eq:partial_sigma_2} into~\cref{eq:ito-Strato-particular}, we obtain
\bal{
    \sigma_t dW_t &= \sigma_t \circ dW_t - \frac{\gamma}{2} \{C^2, \rho_t \} _c dt + 2\gamma m_t \{C, \rho_t\}_c dt - \gamma C\rho_tC^\dagger dt - 4\gamma m_t^2 \rho_t dt + \gamma \braket{\{ C^\H, C \}_c}_t \rho_t dt \\
    & = \sigma_t \circ (2\gamma m_t dt + dW_t) - \frac{\gamma}{2} \{C^2, \rho_t \} _c dt - \gamma C\rho_tC^\dagger dt + \gamma \braket{\{ C^\H, C \}_c}_t \rho_t dt \\
    & = \sigma_t \circ dy_t - \frac{\gamma}{2} \{C^2, \rho_t \} _c dt - \gamma C\rho_tC^\dagger dt + \gamma \braket{\{ C^\H, C \}_c}_t \rho_t dt
    \label{eq:sigma_strato}
}
where, in the second line, we used the definition of $\sigma_t$, and, in the third line, the definition of the measurement process $dy_t = 2\gamma m_t dt + dW_t$.

Finally, inserting~\cref{eq:sigma_strato} into the It\^o SME~\cref{eq:sme_ito_sigma} and expanding the Lindbladian term using the definition of $\cL(\rho_t)$, we obtain the SME in Stratonovich form:
\bal{\label{eq:drho_strato}
    d\rho_t =& - i[H_0, \rho_t] dt - \frac{\gamma}{2} \lpar \{C^\dagger C, \rho_t\} + \{C^2, \rho_t \}_c -2 \braket{\{ C^\H, C \}_c}_t \rho_t \rpar dt + \{ C - m_t, \rho_t \}_c \circ  dy_t\,.
}

\bigskip
In the special case where the jump operator is Hermitian $C= C^\dagger$, the Stratonovich SME reduces to
\bal{\label{eq:sme_hermitian_case}
    d\rho_t = - i[H_0, \rho_t] dt -\gamma \{C^2, \rho_t\} dt + 2\gamma \braket{C^2}_t \rho_t dt + \{C - m_t, \rho_t\} \circ dy_t
}
which matches the monitored dynamics derived in~\cite{garcia-pintosReshapingQuantumArrow2025}.
To see this, define $\gamma = \frac{1}{4\tau}$, with $\tau$ as in~\cite{garcia-pintosReshapingQuantumArrow2025}, and $dW_t = \sqrt{\gamma} dB_t$ with $dB_t^2  = dt$.
The measurement process in~\cite{garcia-pintosReshapingQuantumArrow2025} is defined as $r_t dt = m_t dt + \sqrt{\tau}\, dB_t$ with $m_t = \braket{C^\H}_t = \braket{C}_t$, while in our formalism it is defined as $dy_t = 2\gamma m_t dt + dW_t$.
These definitions are equivalent by setting $dy_t = \frac{r_t}{2 \tau} dt$.

%================================================================================

\subsection{A stochastic Hamiltonian that describes monitored dynamics} \label{app:stoch_ham}

Here we derive the Hamiltonian generator of the SME~\cref{eq:drho_strato}.
As we work in the Stratonovich picture, we omit the $\circ$ notation henceforth.

To identify a Hamiltonian generator of the dynamics we need to write the right-hand side of the SME~\cref{eq:drho_strato} as a commutator with the state $\rho_t$.
For the last term in~\cref{eq:drho_strato}, this is straightforward using the identity $\{C - m_t, \rho_t\}_c = [[A, \rho_t] + B, \rho_t]$~\cref{eq:conj_anticomm} where $A= C^\H$ and $B = C^\AH$.
Thus, we write
\bal{
\{ C - m_t, \rho_t \}_c dy_t = -i[H_M(t), \rho_t]dt
}
where the measurement Hamiltonian $H_M(t)$ is given by
\bal{\label{eq:app_HM}
H_M(t) =& i\lpar [A, \rho_t] + B \rpar \frac{dy_t}{dt}\,.
}

To derive a Hamiltonian generator for the second term in~\cref{eq:drho_strato}, which is proportional to $\gamma$, we use the following identity: for an arbitrary operator $\Omega$ and pure state $\rho_t$, it holds that
\baln{
[[\Omega,\rho_t],\rho_t]
&= \Omega\rho_t^2 - 2\rho_t\Omega\rho_t + \rho_t^2\Omega \\
&= \{\Omega, \rho_t\} - 2 \braket{\Omega}_t \rho_t\,,
}
where, in the second line, we used that $\rho_t^2 = \rho_t$, and $\rho_t \Omega \rho_t = \braket{\Omega}_t \rho_t$.
In particular, for $\Omega = \{A, C\}_c$, where $A = C^\H$, we have
\bal{\label{eq:braket_1}
[[\{A, C\}_c, \rho_t], \rho_t] =& \{\{A,C\}_c,\rho_t\} - 2\braket{\{A, C\}_c}_t \rho_t
}
We expand the first term in the left-hand side of~\cref{eq:braket_1} as
\baln{
\{\{A,C\}_c,\rho_t\}
&=
\frac{1}{2}\{C^2,\rho_t\}
+
\{C^\dagger C,\rho_t\}
+
\frac{1}{2}\{(C^\dagger)^2,\rho_t\} \\
&= \frac{1}{2}\bigl(C^2\rho_t+\rho_t(C^\dagger)^2\bigr)
+
\frac{1}{2}\bigl(\rho_t C^2+(C^\dagger)^2\rho_t\bigr)
+
\{C^\dagger C,\rho_t\}.
}
where, in the first line, we used the definition of $A = (C + C^\dagger)/2$, and, in the second line, we expanded anti-commutators and regrouped.
Using the definitions
\baln{
\{C^2,\rho_t\}_c = C^2\rho_t+\rho_t(C^\dagger)^2,
\qquad
\{\rho_t,C^2\}_c = \rho_t C^2+(C^\dagger)^2\rho_t,
}
we arrive at
\bal{
\{\{A,C\}_c,\rho_t\}
=
\frac{1}{2}\Bigl(\{C^2,\rho_t\}_c+\{\rho_t,C^2\}_c\Bigr)
+
\{C^\dagger C,\rho_t\}.
\label{eq:middle}
}
Inserting back into~\cref{eq:braket_1} we obtain
\baln{
[[\{A,C\}_c,\rho_t],\rho_t]
=
\frac{1}{2}\Bigl(\{C^2,\rho_t\}_c+\{\rho_t,C^2\}_c\Bigr)
+
\{C^\dagger C,\rho_t\}
-
2\langle \{A,C\}_c\rangle_t \rho_t.
}
Rearranging,
\baln{
[[\{A,C\}_c,\rho_t],\rho_t]
-
\frac{1}{2}\Bigl(\{C^2,\rho_t\}_c+\{\rho_t,C^2\}_c\Bigr)
=
\{C^\dagger C,\rho_t\}
-
2\langle \{A,C\}_c\rangle_t \rho_t\,,
}
and add $\{C^2, \rho_t\}_c$ to both sides:
\bal{
\label{eq:braket_2}
[[ \{A,C\}_c, \rho_t ], \rho_t]
- \frac{1}{2}\Bigl( \{\rho_t,C^2\}_c - \{C^2,\rho_t\}_c  \Bigr)
= \{C^2, \rho_t\}_c
+ \{C^\dagger C,\rho_t\}
- 2\langle \{A,C\}_c\rangle_t \rho_t\, .
}
Note that the right-hand side of~\cref{eq:braket_2} is identical to the term between parenthesis multiplying the damping rate $\gamma$ in~\cref{eq:drho_strato}.

Expand
\baln{
\{\rho_t, C^2\}_c - \{C^2, \rho_t\}_c &= [\rho_t, C^2] + [(C^\dagger)^2, \rho_t] \\
&= -[C^2 - (C^\dagger)^2, \rho_t]
}
and replace back in~\cref{eq:braket_2}:
\bal{\label{eq:braket_3}
[[\{A, C\}_c, \rho_t] + \frac{1}{2} \lpar C^2 - (C^\dagger)^2 \rpar, \rho_t] = \{C^2, \rho_t\}_c + \{C^\dagger C, \rho_t\} - 2\braket{\{A, C\}_c} \rho_t
}
The left-hand side of~\cref{eq:braket_3} is a commutator with the state, while the right-hand side corresponds to the $\gamma$ term in~\cref{eq:drho_strato}.
Define the dissipation Hamiltonian
\bal{\label{eq:app_HD}
H_D(t) = -i\frac{\gamma}{2} \lpar [\{A, C\}_c, \rho_t] + \frac{1}{2} \lpar C^2 - (C^\dagger)^2 \rpar \rpar
}

Finally, we collect the Hamiltonians $H_D$~\cref{eq:app_HD} and $H_M$~\cref{eq:app_HM} to write the monitored dynamics \cref{eq:drho_strato} as a stochastic quantum Liouville equation
\bal{\label{eq:dP_stoch_unitary}
    d\rho_t = -i[H_0 + H_D(t) + H_M(t), \rho_t] dt\,.
}

Given that $\rho_t$ is pure, equation~\cref{eq:dP_stoch_unitary} corresponds to a stochastic unitary evolution with propagator in the form of a Dyson-Stratonovich time-ordered exponential
\bal{\label{eq:stoch_prop}
    U = \cT \exp \lkey 
    -i\int_0^T \lpar H_0 + H_D(t) + H_M(t) \rpar dt
    \rkey\,.
}
Note that there are no Itô corrections in the expression above as we are working in the Stratonovich picture, where the standard rules of calculus apply.

To conclude the proof of Theorem~\ref{th:sme_strato}, we write~\cref{eq:dP_stoch_unitary} explicitly in terms of the Hermitian part $A=C^\H$ and anti-Hermitian part $B=C^\AH$.
Expand~\cref{eq:dP_stoch_unitary} using the definitions for $H_D$ and $H_M$:
\bal{\label{eq:drho_strato_explicit}
     d\rho_t = -i[H_0, \rho_t]dt -\frac{\gamma}{2}\lbra \lbra \{A, C\}_c, \rho_t\rbra, \rho_t \rbra dt -\frac{\gamma}{4} \lbra C^2 - (C^\dagger)^2, \rho_t \rbra dt +
     [[A, \rho_t], \rho_t]dy_t + [B, \rho_t]dy_t
}
Compute
\bal{
\{A, C\}_c &= \{A, A + B\}_c \\
            &= 2A^2 + AB - BA \\
            &=2A^2 + [A, B]\,,
            \label{eq:id_1}
}
and
\bal{
C^2 - (C^\dagger)^2 &= (A + B)^2 - (A - B)^2 \\
                    &=2\{A, B\}\,,
                    \label{eq:id_2}
}
where we used that $B^\dagger = (C^\AH)^\dagger = -B$.
Replace~\cref{eq:id_1} and~\cref{eq:id_2} back into~\cref{eq:drho_strato_explicit}:
\bal{\label{eq:final_strato}
     d\rho_t &= -i[H_0, \rho_t]dt
            -\gamma [[A^2, \rho_t], \rho_t]dt
            + [[A, \rho_t], \rho_t] dy_t
            - \frac{\gamma}{2}[[[A, B], \rho_t], \rho_t]dt
            -\frac{\gamma}{2} [ \{A, B\}, \rho_t] dt
            + [B, \rho_t] dy_t \\
            &=  -i\left[ H_0 dt -i\frac{\gamma}{2} \{A, B\} dt
            + i B dy_t,\, \rho_t \right]
            + \left[ \left[ -\gamma A^2 dt - \frac{\gamma}{2}[A, B] dt +  A dy_t, \rho_t\right],\, \rho_t\right]\,.
            \label{eq:braket_4}
}
By identifying in~\cref{eq:braket_4} the processes
\baln{
dH_t^\text{SB} = H_0 dt -i\frac{\gamma}{2} \{A, B\} dt + i B dy_t
}
and
\baln{
dH_t^\text{DB} = -\gamma A^2 dt - \frac{\gamma}{2}[A, B] dt +  A dy_t\,,
}
we arrive at
\baln{
d\rho_t = -i[dH_t^\text{SB},\, \rho_t] + [[dH_t^\text{DB},\, \rho_t],\, \rho_t]\,,
}
and Theorem~\ref{th:sme_strato} is proved.

The generalization to multiple noise channels is straightforward following the definitions of App.~\ref{app:ito}.

%==================================================

\section{Double-bracket gradient flows}\label{app:gradient_flow}

\subsection{Mathematical background}
\label{app:math_back}

It is well known that double-bracket flows can be interpreted as Riemannian gradient flows in the unitary orbit~\cite{HelmkeMoore1994, brockett1993differential, schulte-herbruggenGradientFlowsOptimization2010}.
In order to prove Theorem~\ref{th:grad_flow} in the main text, we provide here a simplified derivation based on~\cite{schulte-herbruggenGradientFlowsOptimization2010}, invoking minimal concepts from differential geometry~\cite{nakaharaGeometryTopologyPhysics2003} and Lie group theory~\cite{rossmannLieGroupsIntroduction2006, knappLieGroupsIntroduction1996}.

The Lie algebra $\mathfrak{u}(N)$ of the unitary group $U(N)$ is given by all anti-Hermitian matrices $X^\dagger = -X$.
Define a Riemannian metric on the unitary group $U(N)$ given by the Hilbert-Schmidt (HS) inner product $\braket{X, Y}_\text{HS} = \tr{X^\dagger Y}$, where $X, Y \in \mathfrak{u}(N)$.

\begin{figure}
    \centering
    \includegraphics[width=0.4\linewidth]{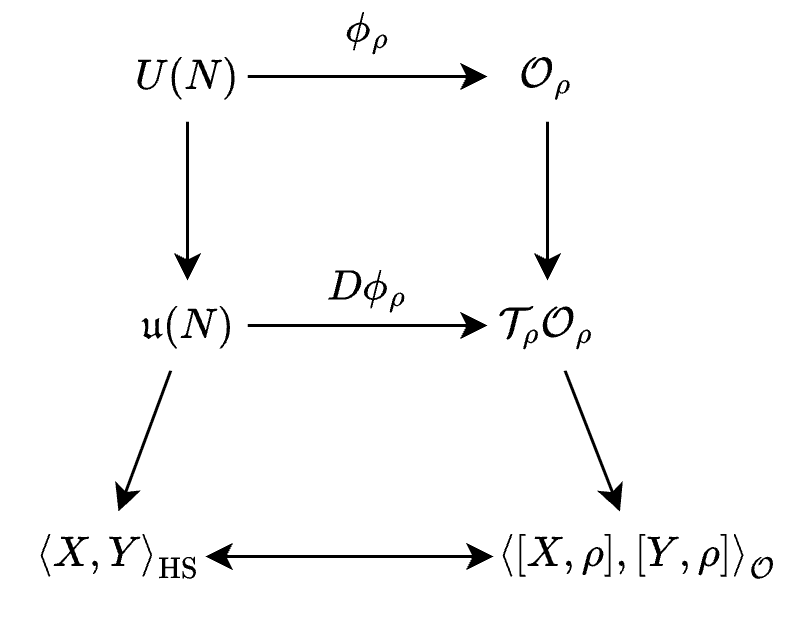}
    \caption{
    The derivative of the action map $\phi_\rho$ maps tangent vectors on the unitary group---in this case, at the identity---to tangent vectors on the orbit of $\rho$.
    This induces a HS metric on the state's orbit such that $\braket{X, Y}_\text{HS} = \braket{[X, \rho], [Y, \rho]}_\cO$.
    }
    \label{fig:inner_prod_diagram}
\end{figure}

Given a state $\rho$, define the action map $\phi_\rho : U(N) \goto \cO_\rho$ such  that $\phi_\rho(U) = U\rho U^\dagger$.
Consider the tangent space to the unitary orbit composed of all commutators of the form $[X, \rho]$ with $X$ in the Lie algebra $\mathfrak{u}(N)$~\cite{rossmannLieGroupsIntroduction2006, knappLieGroupsIntroduction1996}.
The action map $\phi_\rho$, via its differential map $D\phi_\rho$, induces a (normal) metric $\braket{\bullet, \bullet}_\cO$ in the unitary orbit~\cite{nakaharaGeometryTopologyPhysics2003},
such that for two arbitrary tangent vectors $[X, \sigma]$ and $[Y, \sigma]$ at a point $\sigma$ in the orbit, we have $\braket{[X, \sigma], [Y, \sigma]}_\cO = \braket{X, Y}_\text{HS}$.
Strictly speaking, $X,\,Y$ are orthogonal to the null directions of $\sigma$, that is, to the vectors $V$ in the algebra that do not generate changes in the state, i.e., $[V, \sigma]=0$.
This restriction allows identifying one-to-one tangent vectors $X$ on the unitary group with tangent vectors $[X, \sigma]$ on the state's orbit.
See Fig.~\ref{fig:inner_prod_diagram} for a diagrammatic illustration of the various concepts introduced so far.

Consider an arbitrary Hermitian operator $O$.
Define the scalar function $F(\sigma) = \tr{O\sigma}$ where $\sigma$ is an arbitrary state in the unitary orbit of $\rho$.
As a function of $\sigma$, $F(\sigma)$ admits a differential $DF|_\sigma$ acting on the tangent space of the orbit of $\rho$, such that $DF|_\sigma([X, \sigma]) \in \real$ for arbitrary anti-Hermitian $X$ in the Lie algebra.
The induced metric allows us to define a unique representative vector of the differential $DF$ in the tangent space of the orbit.
This defines a Riemannian gradient induced by the HS metric, which we denote by $\nabla F(\sigma)$.
To derive the form of $\nabla F(\sigma)$ we invoke the compatibility condition just described.
Since $\nabla F(\sigma)$ is a tangent vector in the unitary orbit, it has a form $[Z, \sigma]$ for some $Z$ in the Lie algebra.
Then, for an arbitrary $X$ in the Lie algebra we have
\baln{
    \braket{\nabla F(\sigma), [X, \sigma]}_\cO &\equiv DF|_\sigma ([X, \sigma])\nonumber \\
            \braket{Z, X}_\text{HS} &= \tr{O [X, \sigma]} \nonumber\\
            -\tr{Z X} &= \tr{[\sigma, O] X}
}
where in the second line we used the compatibility condition of the induced metric and that $DF|_\sigma([X, \sigma]) = \frac{d}{d\tau} (F(\sigma(\tau)))|_{\tau=0} = \tr{O[X, \sigma]}$, where $\sigma(\tau) \coloneqq e^{\tau X}\sigma e^{-\tau X}$ is the integral curve corresponding to the tangent vector $[X, \sigma]$~\cite{nakaharaGeometryTopologyPhysics2003, rossmannLieGroupsIntroduction2006, knappLieGroupsIntroduction1996}.
In the third line we used the definition of the HS product and the trace property.
Because the last line holds for arbitrary $X$, we must have $Z = [O, \sigma]$.
Thus, we arrive at the final form of the gradient
\bal{\label{eq:app_grad}
\nabla F(\sigma) = [Z, \sigma] = [[O, \sigma], \sigma]\,.
}

Now that we established the main connection between double-brackets and gradient flows, we apply it to prove the results of Section~\ref{sec:DoubleBracket}.

\subsection{Gradient flow formulations}
\label{app:grad_flow_form}

The relation between double-brackets and gradient flows derived in the previous section implies the following.
Suppose a general double-bracket dynamics of the form
\bal{\label{eq-app:db_flow}
d\rho_t = \cC_\rho^2(dO_t) = [[dO_t, \rho_t], \rho_t]\,,
}
where $\cC_\rho$ is the commutator with the state, and $dO_t$ is some Hermitian generator process.
We write its gradient flow formulation as
\bal{\label{eq-app:grad_flow}
d\rho_t = \nabla F_t (\rho_t) dt\,,
}
where $F_t(\sigma) \coloneqq \tr{\dot O_t \sigma}$ with $\dot O_t \coloneqq dO_t/dt$, and $\sigma$ in the state's orbit.

In the main text, we stated that pure states satisfy
\baln{
d\rho_t = [[d\rho_t, \rho_t], \rho_t]\,.
}
Setting $d O_t \equiv d \rho_t$ and using the gradient flow formulation given by~\cref{eq-app:db_flow,eq-app:grad_flow}, it is immediate to write the relation above as a gradient flow
\baln{
\dot \rho_t = \nabla F_t(\rho_t)
}
where the potential $F_t(\sigma) = \tr{\dot \rho_t \sigma}$.

Along the same lines, we prove Theorem~\ref{th:grad_flow} in the main text as follows.
The full dynamics~\cref{eq:sme_strato} can be written as
\baln{
d\rho_t &= \cC_\rho(-i dH_t^\sb) + \cC_\rho^2(dH_t^\db) \nonumber\\
        &= \cC_\rho^2 \lpar
        \cC_\rho(-i dH_t^\sb) + dH_t^\db \rpar \nonumber\\
        &= \nabla F_t(\rho_t) dt\,.
}
In the second line we used the periodic property $\cC_\rho^3 = \cC_\rho$.
The third line is obtained by setting $dO_t \equiv \cC_\rho(-i dH_t^\sb) + dH_t^\db$ and using the gradient flow formalism derived before.
By splitting $F_t$ into single- and double-bracket terms we arrive at the desired result.

\subsection{The evolution equation of the variance $V(t)$}
\label{app:variance}

We derive Eq.~\cref{eq:strato_variance}.
Consider $V(t) = \tr{\Delta A_t^2 \rho_t}$ as defined in the text and compute
\bal{\label{eq-app:var1}
dV(t) = \tr{d\Delta A_t^2 \rho_t + \Delta A_t^2 d\rho_t} = \tr{\Delta A_t^2 d\rho_t}\,,
}
where we used that $\tr{d\Delta A_t^2 \rho_t} = 0$.
The increment $d\rho_t$ is a tangent vector on the state's orbit.
Then, we can equivalently write the \rhs of~\cref{eq-app:var1} as the differential map of $V_t(\sigma) = \tr{\Delta A_t^2 \sigma}$ applied to $d\rho_t$, that is,
\baln{
dV(t) = DV_t(d\rho_t) &= \braket{\nabla V_t(\rho_t), d\rho_t}_\cO \nonumber \\
&= -\gamma \|\nabla V_t(\rho_t)\|_\cO^2 dt + \braket{\nabla V_t(\rho_t), \nabla K_t(\rho_t)}_\cO dW_t\,,
}
where in the last two steps we used the compatibility condition of the metric and the dynamics $d\rho_t = -\gamma \nabla V_t(\rho_t) dt + \nabla K(\rho_t) dW_t$.

%==============================================================================================

\section{A simple proof of Theorem~\ref{th:feedback}}\label{app:simple_proof}

Let $H_0 \coloneqq \sum_j E_j \Pi_j$ with $\Pi_j$ the $j$-th eigen-projector of $H_0$.
From~\cref{eq:sme_feedback}, 
\baln{
    \frac{d}{dt} \langle \Pi_j \rangle &= - 2\gamma \left( \tr{H_0^2 \rho_t \Pi_j} - \tr{H_0^2 \rho_t \Pi_j \rho_t} \right) = -2 \gamma \langle \Pi_j \rangle \left( E_j^2 - \langle H_0^2 \rangle \right).
}
This equation implies that the populations of all levels with energy $E^2 \leq \langle H_0^2 \rangle$ increase in time (which, in turn, leads to a decrease in $\langle H_0^2 \rangle$).
Moreover, the rate at which the population of the minimum energy eigenstate increases is the highest.
Thus, the lowest energy eigenstate is the only stable fixed point of the dynamics. 

Note that if $\braket{\Pi_0} \equiv 0$, we cannot guarantee the exponential convergence to the ground state of $H_0$.
As a consequence, a sufficient condition for exponential convergence is that the initial state has non-zero overlap with the ground state.
This is in agreement with literature~\cite{gluzaDoublebracketQuantumAlgorithms2024}.

%==============================================================================================

\section{Stability analysis of the measurement-based feedback dynamics}\label{app:stability_analysis}

Consider the feedback-induced dynamics~\cref{eq:sme_state_agnostic_fb}, which we reproduce here as
\bal{\label{eq:app_sme_state_agnostic_fb}
d\rho_t = - \gamma \lbra [A^2, \rho_t-Q], \rho_t \rbra dt
            -\frac{\gamma}{2} \lbra [[A, B], \rho_t-Q], \rho_t \rbra dt
         + \lbra [A, \rho_t-Q], \rho_t \rbra dy_t\,.
}
Given the target $Q = \ket{q}\!\bra{q}$, we want to analyze the stability of the fixed-point solution $Q$.
That is, whether small perturbations around the solution tend to converge to the solution or escape its surroundings.
For this, we study the behavior of the overlap $Q_\parallel(t) \coloneqq \tr{Q\rho_t}$ by computing $dQ_\parallel(t) = \tr{Q d\rho_t}$.
Note that we can write
\bal{\label{eq:app_overlap}
 dQ_\parallel(t) = \sum_Z dQ_\parallel^Z(t),
}
where
\baln{
dQ_\parallel^Z(t) = \tr{[[dZ_t, \rho_t-Q], \rho_t]Q} = \tr{[dZ_t, \rho_t][\rho_t, Q]} - \tr{[dZ_t, Q][\rho_t, Q]}\,,
}
for Hermitian $dZ_t \in \lkey -\gamma A^2dt,\,-\frac{\gamma}{2}[A, B]dt,\,A dy_t \rkey$.

For clarity of exposition, we adopt the Dirac bracket notation.
The first trace above is $\tr{[dZ_t, \rho_t][\rho_t, Q]} = \braket{\{dZ_t,Q\}}_t - 2 \braket{dZ_t}_t Q_\parallel$, where $\braket{\cdot }_t$ is the quantum expectation with respect to $\ket{\psi_t}$.
The second trace is $\tr{[dZ_t, Q][\rho_t, Q]} = 2\braket{dZ_t}_q Q_\parallel - \braket{\{dZ_t, Q\}}_t$, where $\braket{\cdot }_q$ is the quantum expectation with respect to $\ket{q}$.
Thus,
\bal{\label{eq:app_dQZ1}
    dQ_\parallel^Z(t) = 2 \braket{\{dZ_t, Q\}}_t - 2 (\braket{dZ_t}_t + \braket{dZ_t}_q) Q_\parallel\,.
}
We can always parametrize $\ket{\psi_t} = \sqrt{Q_\parallel(t)} \ket{q} + \sqrt{Q_\perp(t)} \ket{q_\perp(t)}$, where $Q_\perp \coloneqq 1-Q_\parallel$ and $\ket{q_\perp}$ is an unit vector resulting from projecting $\ket{\psi_t}$ onto the orthogonal space of $\ket{q}$.
When the projection is null, we simply set $\ket{q_\perp} = 0$.
We use this decomposition to write~\cref{eq:app_dQZ1} as
\bal{\label{eq:dQZ2}
    dQ_\parallel^Z(t) =2 Q_\parallel Q_\perp \lpar \braket{dZ_t}_q - \braket{dZ_t}_{q_\perp} + 2 \sqrt{\frac{Q_\perp}{Q_\parallel}} \text{Re} \braket{q | dZ_t |q_\perp} \rpar
}
From this we see, as expected, that $Q_\parallel = 1$ is a fixed point of the dynamics, but also $Q_\parallel= 0$, meaning that when the system enters the orthogonal subspace of $\ket{q}$, it remains there for the rest of the evolution.

To analyze the stability of $Q_\parallel = 1$, consider the system state is subject to a small perturbation around the solution $\ket{q}$.
Thus, we can write  the overlap $Q_\parallel = 1 - \epsilon$ and $Q_\perp = \epsilon$ for some $\epsilon > 0$ small.
Then, we write the linearization in $\epsilon$ of~\cref{eq:app_overlap} as
\bal{\label{eq:linear_overlap}
d\epsilon = 2\gamma \epsilon \lkey
\sum_Z \lpar \braket{Z}_q - \braket{Z}_{q_\perp} \rpar
- 2 \lpar \braket{A}_q - \braket{A}_{q_\perp} \rpar \braket{A}_q
\rkey dt
- 2 \epsilon \lpar \braket{A}_q - \braket{A}_{q_\perp} \rpar dW_t
}
where $Z \in \lkey A^2, \frac{1}{2}[A, B] \rkey$, and we used the measurement output formula $dy_t = 2\gamma \braket{A}_t dt + dW_t$ to write the linearization in terms of the Wiener increment $dW_t$.

Note that $d\epsilon^2 \propto \epsilon^2 dt$, that is, the variance in the overlap tends to vanish close to the boundary $Q_\parallel = 1$.
This means it becomes increasingly harder for the system to leave the surroundings of $\ket{q}$.

Recall that we are dealing with Stratonovich equations.
For this solution to be (stochastically asymptotically) stable~\cite{kloeden2012numerical} the Lyapunov exponent associated to the linear equation~\cref{eq:linear_overlap} must be negative.
This implies that
\bal{\label{eq:app_lyapunov_ineq}
\sum_{Z \in \lkey A^2, \frac{1}{2}[A, B] \rkey}
\lpar \braket{Z}_q - \braket{Z}_{q_\perp} \rpar
- 2 \lpar \braket{A}_q - \braket{A}_{q_\perp} \rpar \braket{A}_q < 0 \,.
}

Given a target $\ket{q}$, we can engineer suitable $A, B$ such that the inequality above is met.
Conversely, if $A, B$ are given, we can study the stability regions of the feedback dynamics~\cref{eq:app_sme_state_agnostic_fb} by solving the inequality~\cref{eq:app_lyapunov_ineq}.

In the case of multiple $n_c$ noise channels the condition above generalizes to
\bal{\label{eq:app_lyapunov_general}
    \sum_{a=1}^{n_c} \gamma_a \lpar \sum_{Z_a} \lpar \braket{Z_a}_q - \braket{Z_a}_{q_\perp} \rpar
- 2 \lpar \braket{A_a}_q - \braket{A_a}_{q_\perp} \rpar \braket{A_a}_q \rpar < 0
}
where $Z_a \in \lkey A_a^2, \frac{1}{2}[A_a, B_a] \rkey$ for the $a$-th noise channel with jump operator $C_a$.

We apply the previous analysis to the single-qubit example of the main text.
The jump operators are $\sigma_+,\, \sigma_-$ with damping rates $\gamma_+,\, \gamma_-$, respectively.
Decompose the jump operators in Hermitian and anti-Hermitian parts as $A_+ = A_- = \frac{\sigma_x}{2}$ and $B_\pm = \pm \frac{i\sigma_y}{2}$.
The inequality~\cref{eq:app_lyapunov_general} implies
\bal{\label{eq:app_delta_gamma_1}
-\frac{\Delta \gamma}{4} (\braket{\sigma_z}_q - \braket{\sigma_z}_{q_\perp}) - (\gamma_+ + \gamma_-) (\braket{\sigma_x}_q - \braket{\sigma_x}_{q_\perp}) \braket{\sigma_x}_q < 0
}
where $\Delta \gamma  = \gamma_+ - \gamma_-$.
Writing the target state in polar form as $\ket{q} = \cos(\frac{\theta}{2}) \ket{0}+ e^{i\phi} \sin(\frac{\theta}{2})\ket{1}$, and the (unique up to a global phase) orthogonal state as $\ket{q_\perp} =  \sin(\frac{\theta}{2}) \ket{0} - e^{i\phi}\cos(\frac{\theta}{2}) \ket{1}$ we compute the corresponding expectations values in~\cref{eq:app_delta_gamma_1}.
From these expressions, we arrive at the desired condition:
\baln{
\Delta \gamma \cos(\theta) + 2 (\gamma_+ +\gamma_-) \sin^2(\theta) \cos^2(\phi) > 0\,.
}

%==============================================================================================

\section{Proof of stability for $\lambda_1 =\lambda_2$}
\label{app:stab_entanglement}

The stability condition is
\bal{\label{eq:app_lyapunov_general}
    \sum_{a=1}^{n_c} \gamma_a \lpar \sum_{Z_a \in \lkey A_a^2, \frac{1}{2}[A_a, B_a] \rkey} \lpar \braket{Z_a}_q - \braket{Z_a}_{q_\perp} \rpar
- 2 \lpar \braket{A_a}_q - \braket{A_a}_{q_\perp} \rpar \braket{A_a}_q \rpar < 0
}

We have two noise channels given by
\bal{
C_1 = X_1 (1 - Z_1Z_2)/2\,,\quad C_2 = Z_1 (1 - X_1 X_2)/2\,,
}
with $\gamma_1$ and $\gamma_2$ are the damping rates for $C_1$ and $C_2$, respectively.
The Hermitian and anti-Hermitian parts are $A_1 = \frac{X_1}{2}$ and $B_1 = \frac{i}{2}Y_1 Z_2$, which gives $[A_1, B_1] = -\frac{1}{2}Z_1 Z_2$, and $A_2 = \frac{Z_1}{2}$ and $B_2 = -\frac{i}{2}Y_1 X_2$, which gives $[A_2, B_2] = -\frac{1}{2}X_1 X_2$.

The target state is $Q_\alpha = \ket{q_\alpha} \bra{q_\alpha}$, where $\ket{q_\alpha}$ is defined in Schmidt form as $\ket{q_\alpha} = \sqrt{\alpha} \ket{00} + \sqrt{1-\alpha}\ket{11}$ for $\alpha \in [0, \, 1]$.

For simplicity, consider $\gamma_1 = \gamma_2 = \gamma$.
Using that $(A_a)^2 \propto 1$ and the commutation relations between $A_a$ and $B_a$, the stability condition becomes
\bal{
\braket{Z_1Z_2}_q - \braket{Z_1Z_2}_{q_\perp}
+ 2 \lpar \braket{X_1}_q - \braket{X_1}_{q_\perp} \rpar \braket{X_1}_q
+ \braket{X_1X_2}_q - \braket{X_1X_2}_{q_\perp}
+ 2 \lpar \braket{Z_1}_q - \braket{Z_1}_{q_\perp} \rpar \braket{Z_1}_q
> 0
}

We use that
\begin{gather}
\braket{Z_1Z_2}_q = 1 \\
\braket{X_1}_q = 0 \\
\braket{X_1X_2}_q = 2\sqrt{\alpha(1-\alpha)} = 2r\\
\braket{Z_1}_q = 2\alpha - 1 = s
\end{gather}
We write the stability condition as
\bal{
1 - \braket{Z_1Z_2}_{q_\perp}
+ 2r - \braket{X_1X_2}_{q_\perp}
+ 2 s \lpar s - \braket{Z_1}_{q_\perp} \rpar
= S(q_\perp, \alpha)
> 0
}

Note that for $\alpha = 1/2$, $r=1/2$ and $s=0$, which gives Eq.~(25) in the paper.

We choose an orthonormal basis for the orthogonal space to $\ket{q_\alpha}$ as
\bal{
\ket{e_1} &= \ket{e_\alpha} = \sqrt{1-\alpha} \ket{00} - \sqrt{\alpha} \ket{11} \\
\ket{e_2} &= \ket{01} \\
\ket{e_3} &= \ket{10}
}
We parametrize the orthogonal state $\ket{q_\perp}$ as
\bal{
\ket{q_\perp} = u \ket{e_\alpha} + v \ket{01} + w \ket{10}
}
where $|u|^2 + |v|^2 + |w|^2 = 1$.
We have,
\begin{gather}
\braket{Z_1Z_2}_{q_\perp} = |u|^2 - |v|^2 - |w|^2 \\
\braket{X_1X_2}_{q_\perp} = -2 r |u|^2 + w^*v + w v^* \\
\braket{Z_1}_{q_\perp} = -s |u|^2 + |v|^2 - |w|^2
\end{gather}

We have,
\bal{
S(q_\perp, \alpha) &=
1 - |u|^2 + |v|^2 + |w|^2 + 2r +2 r |u|^2 - (w^*v + \hc) + 2s(s +s |u|^2 - |v|^2 + |w|^2) \\
&= 2(|v|^2 + |w|^2) + 2r(1+ |u|^2) - (w^*v + \hc) + 2s^2(1+ |u|^2) - 2s|v|^2 + 2s|w|^2 \\
&= 2(r + s^2)(1+ |u|^2) + 2(1-s)|v|^2 + 2(1+s)|w|^2 - (w^*v + \hc)  \label{eq:S1}\\
&= 2(r + s^2)(1+ |u|^2) + z^\dagger A z \label{eq:S2}
}
where $z = (v, w)^T$ and
\bal{
A = \pmat{2(1-s) & -1 \\
          -1 & 2(1+s)
}
}
The eigenvalues are $\lambda_\pm = 2 \pm \sqrt{5 - 4(1-s^2)}$, which implies $\min \lambda_- = 2 - \sqrt{5} < 0$.
To make the eigenvalues more informative, we subtract some terms from the first part of the sum in~\cref{eq:S2} using $|u|^2 + |v|^2 + |w|^2 = 1$.
We have
\bal{
S(q_\perp, \alpha) &=
2(r + s^2)(2|u|^2 + |v|^2 + |w|^2) + 2(1-s)|v|^2 + 2(1+s)|w|^2 - (w^*v + \hc) \\
&= 4(r + s^2)|u|^2 + 2(r + s^2 - s + 1)|v|^2 + 2(r + s^2 + s + 1)|w|^2 - (w^*v + \hc) \\
&= 4(r + s^2)|u|^2 + z^\dagger A z
}
where, now,
\bal{
A = \pmat{
a & -1 \\
-1    & b
}
}
where $a = 2(r + s^2 - s + 1)$ and $b = 2(r + s^2 + s + 1)$.
The eigenvalues are
\bal{
\lambda_\pm &= \frac{a+b \pm \sqrt{(a+b)^2 - 4 (ab-1)}}{2} \\
&= 2(r + s^2 + 1) \pm \sqrt{1 + 4s^2}
}
Using that $\sqrt{1+x} \leq 1+\frac{x}{2}$,
\bal{
\lambda_- \geq 2(r+s^2+1)-(1+2s^2)
=1+2r \geq 1\,,
}
which proves that
\bal{
S(q_\perp, \alpha) > 0
}
and, consequently, establishes the stability of $Q_\alpha$ for all $\alpha \in [0, 1]$.

\end{document}